\documentclass[aps,twocolumn,showpacs,preprintnumbers,nofootinbib,prd,superscriptaddress,10pt]{revtex4-2}

\makeatletter
\def\l@subsubsection#1#2{}
\def\l@subsubsubsection#1#2{}
\makeatother

\usepackage{graphicx,amssymb,amsmath,amsthm,amsfonts,epsfig,epsf,fixmath}
\usepackage[usenames]{color}
\usepackage{epstopdf}
\usepackage{isotope}

\usepackage{aas_macros}
\usepackage{bm}
\usepackage{comment}
\usepackage{dcolumn}
\usepackage{latexsym}
\usepackage{rotating}
\usepackage{longtable}

\usepackage{enumerate}
\usepackage{tensor,multirow}
\usepackage{url}
\usepackage[linktocpage]{hyperref}
\usepackage{soul}

\newcommand{\tn}{\textnormal}

\newcommand{\GSSI}{Gran Sasso Science Institute (GSSI), I-67100 L’Aquila, Italy}
\newcommand{\GranSasso}{INFN, Laboratori Nazionali del Gran Sasso, I-67100 Assergi, Italy}

\begin{document}
\title{Sensitivity of Neutron Star Observations 
to Three-nucleon Forces}

\author{Andrea Sabatucci}
\affiliation{Dipartimento di Fisica, ``Sapienza'' Universit\`a di Roma, Piazzale
Aldo Moro 5, 00185, Roma, Italy}
\address{Sezione INFN Roma1, Roma 00185, Italy}

\author{Omar Benhar}
\affiliation{Dipartimento di Fisica, ``Sapienza'' Universit\`a di Roma, Piazzale
Aldo Moro 5, 00185, Roma, Italy}
\address{Sezione INFN Roma1, Roma 00185, Italy}

\author{Andrea Maselli}
\address{\GSSI}
\address{\GranSasso}

\author{Costantino  Pacilio}
\affiliation{Dipartimento di Fisica, ``Sapienza'' Universit\`a di Roma, Piazzale
Aldo Moro 5, 00185, Roma, Italy}
\address{Sezione INFN Roma1, Roma 00185, Italy}
\begin{abstract}

Astrophysical observations of neutron stars have been widely used to 
infer the properties of the nuclear matter equation of state. 
Beside being a source of information on average properties of 
dense matter, the data provided by 
electromagnetic and gravitational wave (GW) facilities are reaching the 
accuracy needed to constrain, for the first time, the underlying nuclear dynamics. In this work we assess the sensitivity of 
current and future neutron star observations to directly infer the strength 
of repulsive three-nucleon forces, which are key to determine 
the stiffness of the equation of state. Using a Bayesian approach we 
focus on the constraints that can be derived on 
three-body interactions from binary neutron star mergers observed by second 
and third-generation of gravitational wave interferometers. We 
consider both single and multiple observations. For current 
detectors at design sensitivity the analysis suggests that 
only low mass systems, with large signal-to-noise ratios (SNR), allow 
to reliably constrain the three-body forces. 
However, our results show that a single observation with a third-generation 
interferometer, such as the Einstein Telescope or Cosmic Explorer, 
will constrain the strength of the repulsive three-nucleon potential 
with exquisite accuracy, turning third-generation GW detectors 
into new laboratories to investigate the properties of nucleon interactions. 
\end{abstract}

\preprint{ET-0134A-22}

\maketitle

\section{Introduction}

Lying at the interface between electromagnetic (EM) observatories, 
gravitational wave interferometers, and Earth based laboratories, 
multi-messenger astrophysics has the potential to shape a novel 
view of both structure and dynamics of dense nuclear matter.
Mass-radius measurements of rotating pulsars are rapidly 
improving thanks to the information provided by the NASA satellite NICER \cite{Cromartie:2019kug,Fonseca:2021wxt,Riley:2019yda,Miller:2019cac,Riley:2021pdl,Miller:2021qha}, which has recently targeted the most massive neutron 
star (NS) known so far. Remarkably, NICER observations of PSR 
J0030+0451 and PSR J0740+662\textemdash the inferred masses of which are $M= {1.34}_{-0.15}^{+0.16}$ ($M=1.44^{+0.15}_{-0.14}M_\odot$) 
and  $M={2.072}_{-0.066}^{+0.067} \  M_\odot$, respectively\textemdash 
yield comparable values of the stellar radius, pointing 
to a stiff nuclear matter equation of state (EOS) up to densities 
around four times
nuclear density. On the other hand, constraints inferred from binary 
NS mergers detected by the LIGO/Virgo Collaboration, and in particular from the 
landmark discovery of GW170817, \cite{LIGOScientific:2018hze,LIGOScientific:2017vwq,LIGOScientific:2020aai,LIGOScientific:2020aai}, have already  
ruled out some of the stiffest EOSs, which predict large tidal deformabilities, hinting instead to a softer matter content
\cite{Hinderer:2007mb,Damour:2009vw,Binnington:2009bb,Flanagan:2007ix,Vines:2010ca,Vines:2011ud}. 
In addition, astrophysical data are being 
complemented by the information coming from terrestrial experiments, 
such as heavy-ion collisions or the recent measurement of the 
neutron skin thickness of lead, performed at Jefferson Lab by the PREX-II 
Collaboration~\cite{2008PPN....39..286C,Li:2013ola,Russotto:2016ucm,Tsang:2008fd,Danielewicz:2002pu,Brown:2013mga,Zhang:2013wna,PREX:2021umo}.

Posterior distributions inferred from space- and ground-based 
facilities have been widely exploited in a variety of multi-messenger 
analyses, aimed at constraining models of the 
EOS or specific properties of neutron star matter. Examples of this approach include reconstruction of the EOS within 
both phenomenological and non-parametric frameworks, calculations based on microscopic models, and analyses focused on 
features such as the occurrence of phase transitions, or the behavior of the symmetry energy above nuclear density \cite{Annala:2017llu,Margalit:2017dij,Radice:2017lry,Bauswein:2017vtn,Lim:2018bkq,Lim:2020zvx,Most:2018hfd,De:2018uhw,Annala:2019puf,Raaijmakers:2019dks,Miller:2019nzo,Kumar:2019xgp,Kumar:2019xgp,Fasano:2019zwm,Landry:2020vaw,Guven:2020dok,Traversi:2020aaa,Raaijmakers:2021uju,Zimmerman:2020eho,Silva:2020acr,Sabatucci:2020xwt,Blaschke:2020qqj,Tang:2020koz,Biswas:2020puz,Pacilio:2021jmq,Malik:2022jqc,Altiparmak:2022bke,Gupta:2022qgg}; 
for recent reviews, see also Refs.~\cite{Baiotti:2019sew,Chatziioannou:2020pqz} 
and references therein.

Recently, some of the authors of this article have proposed a 
novel approach, aimed at pushing the analyses based on 
multimessenger astrophysical information to a deeper level 
\cite{Maselli:2020uol}. They argued that the accuracy of the 
currently available data\textemdash as well as that expected 
to be achieved by operating the existing detectors at design sensitivity\textemdash  offer an unprecedented opportunity 
to constrain the microscopic models of nuclear dynamics at 
supranuclear density. The results reported in Ref.~\cite{Maselli:2020uol} 
show that the data set comprising the GW observation of the binary
NS event GW170817, the spectroscopic observation of the millisecond pulsars PSR J0030+0451 performed
by the NICER satellite, and the high-precision measurement of the 
radio pulsars timing of the binary PSR J0740+6620, providing information on 
the maximum NS mass, can, in fact, be exploited to infer quantitative insight 
on the strength of repulsive three-nucleon interactions in dense matter.

Unlike the nucleon-nucleon potential, the models of irreducible three-nucleon 
interactions are totally unconstrained beyond nuclear density. In most 
models, e.g. the Urbana IX potential employed to derive the EOS of Akmal, 
Pandharipande and Ravenhall (APR)~\cite{APR}, the strength of the 
isoscalar repulsive term\textemdash which plays a pivotal role in 
determining the stiffness of the nuclear matter EOS in the region 
relevant to neutron stars\textemdash is determined in such a way as to 
reproduce the empirical equilibrium density 
of isospin-symmetric matter~\cite{UIX_1,UIX_2}. In this context, the 
availability of additional information constraining the three-nucleon 
potential at larger density would be a major breakthrough. 

The present work can be seen as a complementary follow up to the 
pioneering 
study of Ref.~\cite{Maselli:2020uol}. The analysis is first 
extended to consider a near-future scenario, using current interferometers 
at design sensitivity and stacking multiple binary NS observations 
characterised by different masses and distances. In addition, we apply, for 
the first time, the Bayesian approach to gauge the sensitivity of the Einstein 
Telescope (ET), a proposed third-generation ground-based GW observatory \cite{Punturo:2010zz,Hild:2010id,Maggiore:2019uih}

The body of the article is structured as follows. In Sect.~\ref{sec:eos} we 
outline the dynamical model underlying our study, as well as the simple 
parametrisation adopted to characterise the strength of the repulsive 
component of the three-nucleon potential. The datasets considered in the 
analysis and the details of numerical simulations are described in 
Sections~\ref{sec:datasets} and~\ref{sec:sim}, respectively, while the 
results are reported and discussed in Sect.~\ref{sec:results}. Finally, a 
summary of our findings and the prospects for future developments can be 
found in in Sect.~\ref{sec:summary}.

\section{Modelling Nuclear Dynamics Beyond Nuclear Density}\label{sec:eos}
The EOSs considered in our study have been derived using the formalism of non-relativistic nuclear many-body theory (NMBT). 
Within this framework, nuclear matter is pictured as a uniform system of point like nucleons, the dynamics of which is completely determined by the Hamiltonian\footnote{Unless explicitly stated otherwise, we shall use a the system of units in which $\hbar=G=c=1$.}
\begin{equation}
    {H}=\sum_i \frac{p_i^2}{2m}+ \sum_{i<j}v_{ij}+\sum_{i<j<k}V_{ijk}\ ,\label{math:hamil}
\end{equation}
where $m$ and $p_i$ denote the mass and momentum of the $i$-th nucleon, respectively. Interactions between matter constituents are driven by the  nucleon-nucleon (NN) potential $v_{ij}$\textemdash providing an accurate description of the two-nucleon system in both bound and scattering states\textemdash supplemented by the three-nucleon (NNN) potential $V_{ijk}$, whose inclusion is needed to implicitly take into account the occurrence of processes involving the internal structure of the nucleon. As a consequence,  the role of NNN interactions is expected to become more and more important with increasing density.

Starting from Eq.~\eqref{math:hamil}, a number of different EOSs have been obtained  using both different Hamiltonian models and different many-body 
techniques to calculate the ground state energy of nuclear matter as a function of baryon density. Purely phenomenological Hamiltonians, fitted to the properties of two- and three-nucleon systems, 
have been shown to provide a remarkably accurate account of the energies of the ground and low-lying excited states of nuclei with 
mass number $A \leq 12$,  as well as of their radii~\cite{QMC}. In addition, 
they allow to reproduce the empirical value of the equilibrium density of isospin-symmetric matter (SNM); see, e.g., Ref.~\cite{APR}

Over the past two decades, a great deal of attention has been given to a novel generation of nuclear Hamiltonians, derived using the formalism of Chiral Effective Field 
Theory ($\chi$EFT).  Within $\chi$EFT, the nuclear potentials are obtained from effective Lagrangians comprising pion and nucleon degrees of 
freedom, constrained by the chiral symmetry of strong interactions. The main advantage of this approach is the capability to determine two- and many-nucleon 
potentials in a fully consistent fashion. However, being based on a low momentum expansion its applicability is inherently limited to 
densities $\lesssim 2\varrho_0$, with $\varrho_0 = 0.16 \ {\rm fm}^{-3}$ being the saturation density of SNM~\cite{Benhar:IJMPE,Essick2020}.

In this study, we have considered purely phenomenological Hamiltonians, which are expected to be best suited to describe the properties of nuclear matter in the density region 
extending up to $\sim 5 \varrho_0$, relevant to NS applications. The reference line of our analysis is the Hamiltonian comprising the Argonne $v_{18}$ NN potential~\cite{AV18} (AV18) and the 
Urbana IX NNN potential~\cite{UIX_2,UIX_1} (UIX), which has been employed 
to obtain the APR EOS~\cite{Akmal:1997,APR}.

The AV18 potential is written as a sum of eighteen terms, needed to describe the complex operator structure of nuclear forces. It provides an accurate fit of the NN scattering phase-shifts for laboratory-frame energies up to $\sim 600$ MeV, a value typical of NN collisions in 
strongly degenerate matter at density $\varrho \sim 4\varrho_0$~\cite{Benhar:IJMPE}. A comparison with the central densities obtained from the solution of the 
Tolman-Oppenheimer-Volkoff equations~\cite{T,OV} with the APR EOS~\cite{Sabatucci2020} suggests that this phenomenological potential is adequate to describe NSs having  masses as large as $\sim 2.1$ M$_\odot$.

The UIX model  of the NNN interaction is written as the sum of an attractive potential first derived by Fujita and Miyazawa~\cite{Fujita}\textemdash 
describing two-pion exchange NNN processes with excitation of  a $\Delta$-resonance in the intermediate state\textemdash  and a phenomenological repulsive potential; 
the resulting expression is
\begin{equation}
  V_{ijk}=V_{ijk}^{2\pi}+V_{ijk}^R \ .
\end{equation}
The strength of the two-pion exchange contribution is adjusted to reproduce the observed ground state energies of \isotope[3][]{H} and \isotope[4][]{He}, obtained 
from accurate Monte Carlo calculations~\cite{UIX_1}, whereas that of the isoscalar repulsive term is fixed to obtain the empirical saturation density of SNM\textemdash 
inferred from nuclear data\textemdash from variational calculations carried out using advanced many-body techniques~\cite{UIX_2}. 

It should be kept in mind  that the repulsive term $V_{ijk}^R$ implicitly takes into account relativistic corrections to the 
phenomenological two-nucleon potential $v_{ij}$, which  is determined by fitting NN scattering data in the center-of-mass reference frame. In the presence of the nuclear medium, however, the center of mass of the interacting 
nucleon pair is not at rest, and $v_{ij}$ must be boosted to take into account its motion~\cite{boost}.

The authors of Ref.~\cite{APR} have modified the free-space AV18 potential to include the boost correction $\delta v$, whose effect is an  enhancement of the repulsive contribution to the potential energy.  As a consequence, using the boosted AV18 potential in
calculations of nuclear matter energy entails the introduction of a modified NNN potential, referred to as  UIX$^*$, 
which turns out to be considerably softer than the UIX.
The impact of relativistic corrections to the nuclear Hamiltonian on the description of NS properties has been recently discussed in Ref.~\cite{Sabatucci2020}. 

The potentials describing NNN interactions are only determined by nuclear phenomenology reflecting nucleon  
interactions at SNM saturation density. On the other hand, they are totally unconstrained in the high-density 
regime relevant to NSs, in which their contribution is known to become dominant. 

Motivated by the above  consideration, in this work we extend the study of Ref.~\cite{Maselli:2020uol},  whose authors have explored the possibility 
of inferring the strength of the repulsive term of the UIX$^*$ potential from data collected by multimessenger astrophysical observations, 
which carry information on nuclear dynamics at supranuclear denisity. Note that to pin down the dynamics of NNN interactions it is essential that the analysis be carried out 
using the the boost corrected NN potential. 

Our study is based on the use of a set of Hamiltonians, obtained from the AV18 + $\delta v$ + UIX$^*$ model performing the replacement 
\begin{equation}\label{3bodya}
           \langle V_{ijk}^R\rangle\rightarrow \alpha\langle V_{ijk}^R\rangle \ .
\end{equation}
 
The energy-density of nuclear matter at arbitrary baryon density $\varrho$ and 
proton fraction $x_p$ has been obtained generalising  the 
parametrisation employed in Ref.\cite{APR}, that can be written in the form
\begin{align}
\label{energy_fit1}
\epsilon(\varrho,x_p) &=\left[\frac{\hbar^2}{2m}+f(\varrho,x_p)\right]\tau_p \\ 
\nonumber
&+\left[\frac{\hbar^2}{2m}+f(\varrho,1-x_p)\right]\tau_n+g(\varrho,x_p),
\end{align}
where
\begin{equation}
\label{energy_fit2}
g(\rho,x_p)=g(\rho,1/2)+\left[g(\rho,0)-g(\varrho,1/2)\right](1-2x_p)^2.
\end{equation}

The explicit expressions of the functions appearing in Eqs.~\eqref{energy_fit1} and~\eqref{energy_fit2} can be found in the Appendix. They involve a set of parameters which were determined by fitting the energy per nucleon of SNM and pure neutron matter (PNM) computed within the FHNC/SOC variational approach~\cite{bob_vijay_rmp} using the AV18+ $\delta v$ + UIX$^*$ Hamiltonian.

The first two terms of Eq.~\eqref{energy_fit1} correspond to the proton and neutron  kinetic energy, respectively, whereas 
the function $g(\rho,x_p)$ describes the contribution arising from interactions.
The assumption of quadratic dependence of the interaction energy on the neutron excess $\delta = 1 - 2x_p$
is routinely employed in the literature to obtain the EOS of $\beta$-stable matter from those of SNM and PNM, and has been shown to 
be remarkably accurate over a broad range of values of the proton fraction $x_p$; see, e.g. Ref.~\cite{BL:2017}.

Implementing the substitution of Eq.~\eqref{3bodya} is equivalent to adding a term $(\alpha-1)V^R$  at first order in perturbation theory.
The corresponding change of  energy density turns out to be
 \begin{equation}
 g(\varrho,x_p)\rightarrow g(\varrho,x_p,\alpha)=g(\varrho,x_p)+\delta g(\varrho,x_p,\alpha),
 \end{equation}
 with
\begin{align}
\delta g(\varrho,x_p,\alpha) & = \delta g(\varrho,1/2,\alpha)\left[1-(1-2x_p)^2\right] \\ 
\nonumber
  & +  \delta g(\varrho,0,\alpha)(1-2x_p)^2 \ .
\end{align}
The functions $\delta g $ can be readily expressed in terms of expectation values of $V^{R}$ in the nuclear matter ground state using
\begin{align}\label{gVR}
 \delta g(\varrho,1/2,\alpha) = \frac{\varrho}{A}  (\alpha - 1) \langle V^R_{ijk} \rangle_\tn{SNM} \ , \\\label{gVR2}
 \delta g(\varrho,0,\alpha) = \frac{\varrho}{A}  (\alpha - 1) \langle V^R_{ijk} \rangle_\tn{PNM} \ .
 \end{align}

Tabulated values of $\langle V_{ijk}^R\rangle$ as a function of density can be found in Ref.~\cite{APR}.
In our analysis, we have employed a polynomial fit including powers up to $\varrho^3$
\begin{equation}
\label{3BF:fit}
\langle V^R_{ijk}\rangle=a_0+a_1\,\varrho+a_2\,\varrho^2+a_3\,\varrho^3 \ ,
\end{equation} 
which turned out to be very accurate. The values of the parameters $a_i$ are reported in Table~\ref{tab:my_label}.
\begin{table*}[hbt]
\caption{Values of the  parameters appearing in Eq.~\eqref{3BF:fit}, corresponding to $\langle V^R_{ijk}\rangle$ in MeV and $\varrho$ in fm$^{-3}$.}
    \label{tab:my_label}
    \begin{tabular}{ccccc}
    \hline\hline
      &  $a_0$  & $a_1$  & $a_2$    & $a_3$    \\
      & $[{\rm MeV}]$ & $[{\rm MeV} \ {\rm fm}^3]$ & $[{\rm MeV} \ {\rm fm}^6]$ & $[{\rm MeV} \ {\rm fm}^9]$   \\ 
      \noalign{\smallskip}  \hline
       SNM  & 0.754  & -16.769  & 214.164    & 77.422  \\
       PNM &  0.949 & -27.403  & 241.407   & 64.995   \\
      \hline\hline
    \end{tabular}
\end{table*}

 Using the analytic expression of the energy density of nuclear matter at arbitrary proton fraction, composition and energy density of $\beta$-stable matter can be easily determined, by minimising with respect to $x_p$, with the additional constraints of conservation of baryon number and charge neutrality. Finally, the matter pressure $P$, derived from standard thermodynamic relations, is used to obtain the EOS $P(\epsilon)$.

It has to be kept in mind that changing the strength of $V^R_{ijk}$ affects the value of the nuclear saturation density predicted by the AV18 + $\delta v$ + UIX$^*$ Hamiltonian. 
For this reason, we have limited the acceptable range of $\alpha$ to the interval $[0.7,2.0]$. Within this range, the departure from the empirical 
value of $\varrho_0$ turns out to be $\sim 15\%$ at most, and the corresponding change of the energy per particle never exceeds 3\%.

Moreover, because the contribution of the repulsive NNN potential becomes large at supranuclear densities, the modification of its strength $\alpha$ marginally affect 
the ground-state energy of atomic nuclei. Using the results reported in Ref.~\cite{Carlson:QMC}, obtained from accurate Quantum Monte Carlo calculations, 
we have found that changing $\alpha$ from 1 to 1.3 results in a change of $4\%$ and $6\%$ of the ground state energies of \isotope[4][]{He} and \isotope[12][]{C}, respectively. 
These discrepancies appear to be fully acceptable in the context of our exploratory study.

\begin{figure}[htbp!]
    \centering
    \includegraphics[scale = 0.55]{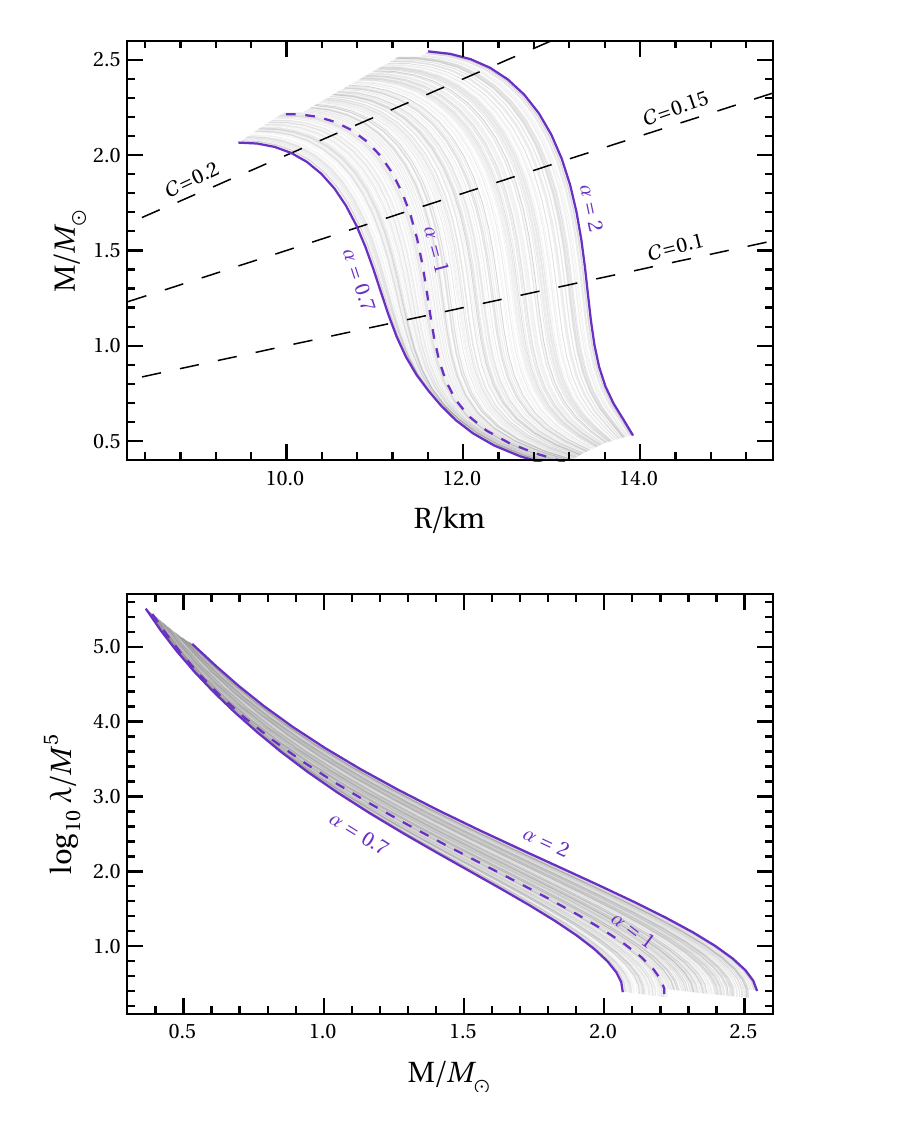}
    \caption{(Top) Representative ensemble of the mass-radius 
    profiles for the family of EOS considered in this work. 
    Each gray curve corresponds to a specific value of 
    $\alpha$ drawn between the solid violet lines which refer 
    to the lower and upper bounds of $\alpha$ assumed in the 
    analysis, i.e. $\alpha=0.7$ and $\alpha=2$, respectively. 
    The dashed curve identifies the baseline APR model 
    with $\alpha=1$. We also show lines of constant compactness 
    ${\cal C}=M/R$. (Bottom) Same as top panel but for the 
    dimensionless tidal deformability $\lambda/M^5$ as a 
    function of the NS mass.}
    \label{fig:eos_profiles}
\end{figure}

\section{Methods and observations}
\label{sec:data}
We consider a family of EOS for which the observables of  a neutron star 
(mass, radius and tidal deformability) depend uniquely on the three-body 
coefficient $\alpha$ and on the central pressure $p_c$:
\begin{equation}
\label{map:eos}
\{\alpha,p_c\} \rightarrow \{M,R,\lambda\}\ .
\end{equation}
Figure~\ref{fig:eos_profiles} shows the stable stellar configurations 
in the mass-radius plane and the mass-tidal deformability plane. Given 
a set $O_{i=1,\dots,n}$ of observations, we infer 
$\{\alpha,p_c^{(1)}\dots p_c^{(m)}\}$ 
\footnote{In general $m\neq n$: for binary coalescence events, we must sample 
over the pressures of both members of the binary.} using a hierarchical 
Bayesian approach,
\begin{equation}
    \label{math:bayes}
    \mathcal{P}(\alpha,\vec{p}_c|\vec{O})\propto\mathcal{P}_0(\alpha,\vec{p}_c)\prod_{i=1}^m\mathcal{L}(O_i|\theta_i)
\end{equation}
where $\vec{p}_c=\{p_c^{(1)}\dots p_c^{(m)}\}$, $\mathcal{L}(O_i|\theta_i)$ is the likelihood of the $i$-th event (see Sec.~\ref{sec:datasets} below) and $\theta_i$ denotes the set of relevant NS observables --- mass and radius for pulsars, symmetric mass ratio and effective tidal deformability for GW observations --- evaluated at $\{\alpha,p_c^{(i)}\}$ via \eqref{map:eos}.
We assume that the priors on $\alpha$ and on each central 
pressure in Eq.~\eqref{math:bayes} are uncorrelated.

The posteriors in Eq.~\eqref{math:bayes} are sampled using the \texttt{emcee} with stretch move \cite{2013PASP..125..306F}. For each observation we run 100 walkers of $10^6$ samples with a thinning factor of $0.02$. The final distribution for $\alpha$ is obtained by marginalizing over the central pressures $\vec{p}_c$. When presenting results, we quote the median alongside the bounds of the $90\%$ symmetric posterior density intervals.

We sample the central pressures of each 
star uniformly in log-space between 
$\ln_{10}p_c^\tn{min}(\alpha)\simeq34.58$, where $p_c$ is expressed in $\mbox{dyne}/\mbox{cm}^2$, and $\ln_{10}p_c^\tn{max}(\alpha)$, 
where $p_c^\tn{max}$ 
corresponds to the central pressure of the heaviest 
stable configuration for each EOS specified by $\alpha$. 
The lower value $p_c^\tn{min}$ is chosen such that 
the nuclear model supports masses larger than $0.8M_\odot$. 
The values of $\alpha$ are drawn from a uniform 
distribution in the range $[0.7,2]$. 
We also impose a causality constraint, 
requiring that the speed of sound $c_s=\sqrt{dp/d\epsilon}$ 
is subluminal at the center of each NS.

\subsection{Astrophysical datasets}
\label{sec:datasets}

We consider three real datasets corresponding to (i) the binary 
coalescence GW170817, (ii) the millisecond pulsar PSR J0030+0451 
and (iii) the heaviest NS observed so far PSR J0740+6620. 
Dataset (iii) provides and update w.r.t. \cite{Maselli:2020uol}, 
in which PSR J0740+6620 was included only through 
the measurement of its mass, while here we also 
include the radius.
We briefly summarize 
here the basic properties of each dataset and the 
corresponding likelihood functions that enter Eq.~\eqref{math:bayes}.

(i) --- GW170817 is the first binary neutron star system 
observed by LIGO and Virgo. 
Under a low spin prior, the LVC analysis constrained the source 
component masses $(m_{1},m_{2})$ between $\sim1.16M_\odot$ and 
$\sim1.6M_\odot$. GW170817 provided the first evidence that 
GW signals from coalescing systems are sensitive to matter 
effects induced by the NS structure, yielding a measurement for 
the effective tidal parameter 
\begin{equation}
\tilde{{\Lambda}}=\frac{16}{13}\left[\frac{(m_1+12m_2)m_1^4\Lambda_1}{(m_1+m_2)^5}\ +\ 1\leftrightarrow2\right]
\end{equation}
of $\tilde{\Lambda}=300^{+420}_{-230}$
within $90\%$ of the highest posterior density interval, 
with $\Lambda_{1,2}=\lambda_{1,2}/m_{1,2}^5$ being the NS individual, dimensionless,  
tidal deformabilities \cite{LIGOScientific:2018hze}.

We construct the likelihood $\mathcal{L}(O_{\rm GW170817}|\eta,\tilde{\Lambda})$ 
from the joint posterior 
${\cal P}({\cal M},\eta,\tilde{{\Lambda}}\vert O_\tn{GW170817})$ 
for $\tilde{\Lambda}$, the chirp mass 
${\cal M}=(m_1 m_2)^{3/5}/(m_1+m_2)^{1/5}$, 
and the symmetric mass ratio $\eta=m_1m_2/(m_1+m_2)^5$.
The calculation can be simplified by the fact that the chirp mass
in the source frame is measured
with $\sim 0.1\%$ precision, 
which allows to fix it to its median value
$\mathcal{M}_\star=1.186~M_\odot$ and restrict to the conditional
probability $\mathcal{P}(\eta,\tilde{\Lambda}|\mathcal{M}_\star,O_{\rm GW170817})$.
Moreover, as shown in \cite{Raaijmakers:2019dks}, 
the latter can be replaced by the marginalized posterior 
${\cal P}(\eta,\tilde{\Lambda}\vert O_\tn{GW170817})$
to very good accuracy. This choice reduces 
the number of parameters to be sampled, since the central
pressure $p_c^{(2)}$ of the secondary component is uniquely
determined by $\{\mathcal{M}_\star,p_c^{(1)}\}$ and $\alpha$
\footnote{More specifically, we compute $m_2$ from $m_1(\alpha,p_c^{(1)})$ and 
$\mathcal{M}_\star$ and then we solve $m_2\equiv m_2(\alpha,p_c^{(2)})$ for $p_c^{(2)}$.}, and similarly 
for the  individual masses $m_{1,2}$ and tidal deformabilities
$\Lambda_{1,2}$.
The likelihood function\footnote{Note that the likelihood we 
use here for GW170817 is different from the one of 
Ref.~\cite{Maselli:2020uol} in which a three-dimensional 
distribution ${\cal L}_\tn{GW}(q,\Lambda_1,\Lambda_2)$ was 
considered, with $q=m_1/m_2$.} is then obtained by re-weighting 
the posterior by the joint prior on $\eta$ and $\tilde{\Lambda}$ 
as derived from \cite{LIGOScientific:2018hze},
\begin{equation}
{\cal L}(O_\tn{GW170817}\vert \eta,\tilde{\Lambda})=\frac{{\cal P}(\eta,\tilde{\Lambda}\vert O_\tn{GW170817})}{{\cal P}_0(\eta,\tilde{\Lambda})}\ .
\label{math:like_lvc}
\end{equation}
Note that, although $p_c^{(2)}$ is not independently sampled, we still
require it to lie within its prior support.

(ii) --- For the millisecond pulsar PSR J0030+0451 we use 
the joint mass-radius posterior ${\cal P}(M,R\vert O_{J0030})$ 
inferred by the NICER collaboration, which has carried 
out two independent studies of the stellar 
spectroscopic observations, obtaining consistent results. 
The mass-radius constraints provided by the two 
collaborations led to 
$M=1.34^{+0.15}_{-0.16}M_\odot$ and 
$R=12.71^{+1.14}_{-1.19}$km \cite{Riley:2019yda}, and 
$M=1.44^{+0.15}_{-0.14}M_\odot$ and 
$R=13.02^{+1.24}_{-1.06}$km \cite{Miller:2019cac} respectively ($68\%$ credibility).
Here we use the data publicly available 
in \cite{riley_thomas_e_2020_5506838},
for which the 
likelihood can be derived straightforwardly 
from ${\cal P}(M,R\vert O_{J0030})$ because the joint 
prior on $\{M,R\}$ is flat,
\begin{equation}
    \mathcal{L}(O_{\rm J0030}|M,R)\propto 
    \mathcal{P}(M,R|O_{\rm J0030})\,.
\end{equation}

(iii) --- PSR J0740+6620 \cite{Riley:2021pdl,Miller:2021qha} 
is the most massive pulsar discovered so far.
Previous observations of this source constrained 
its mass to $M=2.08^{+0.072}_{-0.069}M_\odot$ 
(68.3\% credibility) \cite{Fonseca:2021wxt}. This 
measurement, combined with data obtained from the XMM Newton 
European Photon Imaging Camera to improve 
the NICER background, was used in \cite{Riley:2021pdl,riley_thomas_e_2021_4697625} 
and \cite{Miller:2021qha,miller_m_c_2021_4670689} 
to infer the pulsar radius, 
with the two teams obtaining $R=12.39^{+1.30}_{-0.98}$km 
and $R=13.7^{+2.62}_{-1.50}$km 
\cite{Miller:2021qha} respectively (68\% credibility).
Here we use the data in \cite{raaijmakers_g_2021_4696232}, 
for which the likelihood can be immediately
inferred from the posterior due to uniform priors,
\begin{equation}
    \mathcal{L}(O_{\rm J0740}|M,R)\propto 
    \mathcal{P}(M,R|O_{\rm J0740})\,.
\end{equation}

\subsection{Simulations for 2G and 3G detectors}\label{sec:sim}
We simulate\footnote{We limit our catalogue to 30 events 
because the recovery of the EOS is expected to be biased by
a mismodelling of the underlying BNS population distribution
if the number of sources exceeds $\sim30$ \cite{Wysocki:2020myz}.}
30 binary neutron 
star events for two choices of the three-body strength, 
$\alpha=1$ and $\alpha=1.3$, either for a 
network (HLV) composed by the LIGO Hanford, LIGO Livingston, 
and Virgo detectors at design sensitivity~\cite{Pitkin:2011yk}, 
or for the future third-generation interferometer Einstein 
Telescope in its ET-D configuration~\cite{Hild:2010id}. We inject 
64-second long waveforms into a zero-noise configuration as described 
in~\cite{Wade:2014vqa}, with sky location and inclination uniformily 
distributed over the sky. Posterior parameters are recovered using the \texttt{bilby} software \cite{Ashton:2018jfp, Romero-Shaw:2020owr} for GW injections and parameter estimation. For both injection and recovery, we model binary neutron star signals with the \texttt{IMRPhenomPv2\_NRTidal} waveform template~\cite{Dietrich:2017aum,Dietrich:2018uni}. Injected binaries are nonspinning, while component spins are recovered 
imposing a low-spin prior $\chi_{1,2}\in[-0.05,0.05]$ and assuming that spins are (anti-) aligned. 

We assume that tidal parameters are recovered 
uniformly w.r.t.~$\tilde{\Lambda}$ and the tidal parameter
$\delta\Lambda$ which contributes at higher post-Newtonian order in the waveform phase expansion \cite{Castro:2022mpw}, with the 
additional constraint that the individual deformabilities 
$\Lambda_{1,2}$ of the binary components lie between 
$0$ and $5000$.

\begin{figure}[htbp!]
    \centering
    \includegraphics[scale = 0.44]{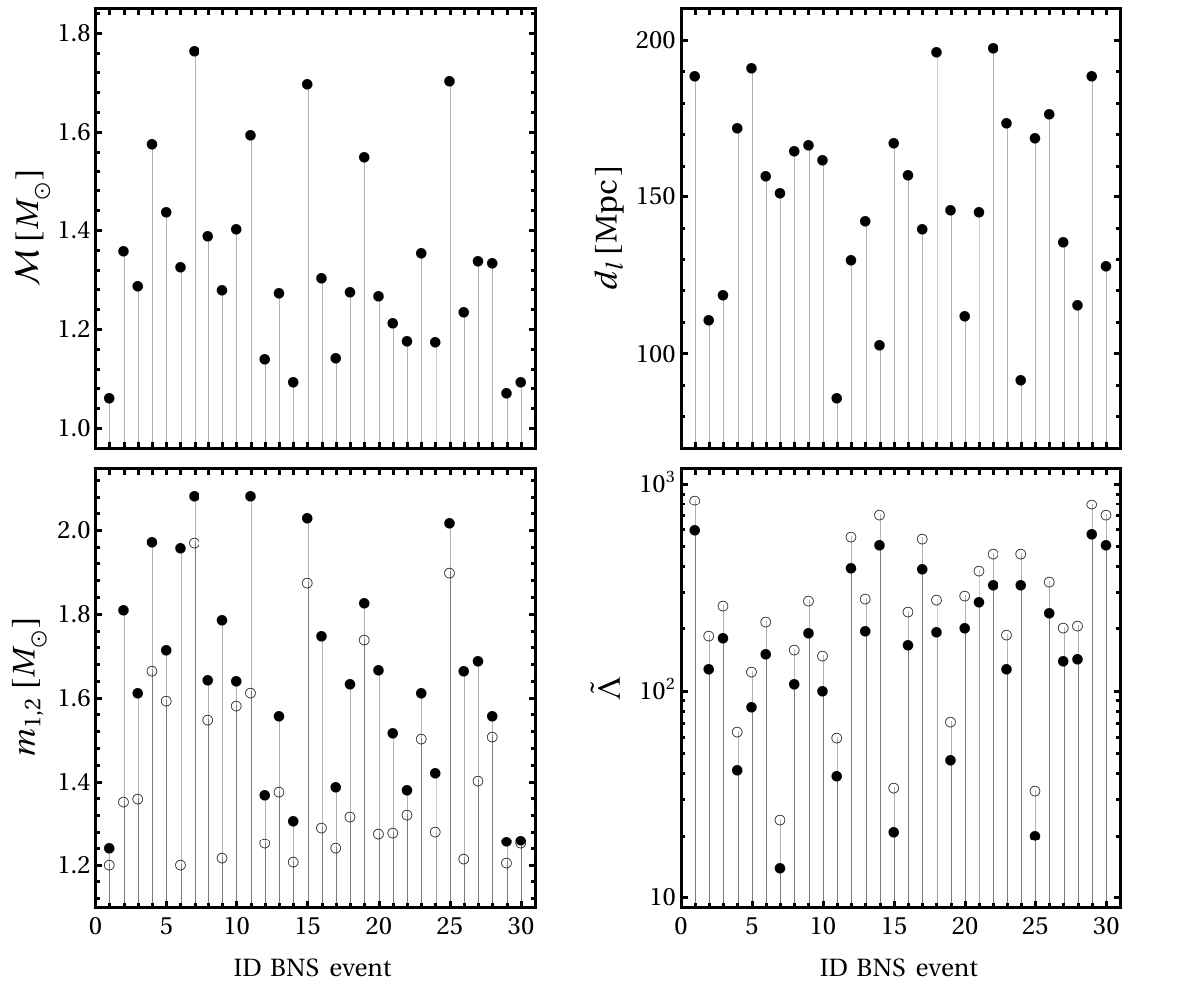}
    \caption{Component masses, luminosity distance, 
    chirp mass, and tidal parameter for the 
    catalogue of NS binaries simulated for HLV 
    and ET observations. Full and empty dots in 
    the left bottom panel correspond to values 
    of $m_1$ and $m_2$, with $m_1 \geq m_2$. Full 
    and empty markers in the bottom right plot 
    identify the tidal parameter for the two 
    values of $\alpha$ we considered, $\alpha=1$ 
    and $\alpha=1.3$, respectively.}\label{fig:injections}
\end{figure}

\section{Results}
\label{sec:results}

We start the discussion of our results by focusing 
first on the the Bayesian analysis applied to the three 
real observations described in the previous section. 

\begin{figure}[htbp!]
    \centering
    \includegraphics[scale = 0.35]{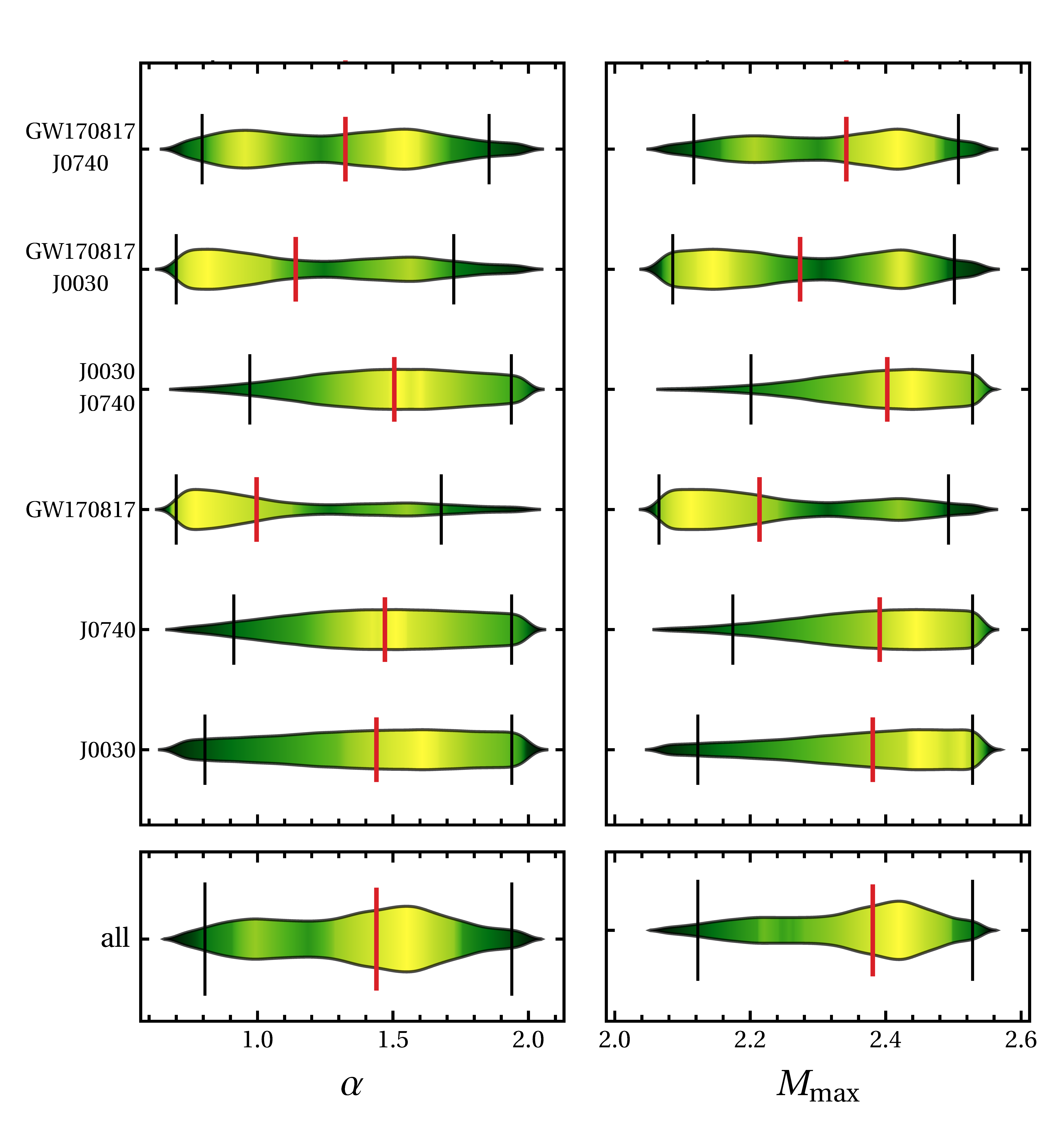}
    \caption{(Left Row) Posterior probability 
    densities for the three-body strength 
    $\alpha$ inferred from different astrophysical 
    datasets. (Right row) 
    Posterior densities for the maximum 
    mass allowed by the EOS corresponding to the 
    inferred distribution of $\alpha$. Bottom 
    panels provide 
    results with all datasets stacked together. 
    Vertical red and black lines identify the 
    median and the the 90\%  posterior 
    density intervals of each distribution, 
    respectively.}
    \label{fig:3BC_real}
\end{figure}

The inferred probability distributions for $\alpha$ 
are summarized by the density plots in the left 
column of Fig.~\ref{fig:3BC_real}, together with 
their median values and 90\% confidence intervals. 
The analyses for GW170817 and for J0030+0451 have 
been already presented in \cite{Maselli:2020uol}, while the novel 
mass-radius measurement obtained by NICER allows us 
to perform an independent study of the three-body 
strength for J0740+6620, and a direct comparison 
with other observations. 
Interestingly the posterior densities of 
Fig.~\eqref{fig:3BC_real} show very similar results for 
the two EM observations, with a nearly identical median 
around $\alpha\simeq1.4$. The probability 
distribution for J0740+6620 peaks around a slightly 
larger value compared to the lighter pulsar, 
J0030+0451, since larger values of $\alpha$ tend to 
support more massive configurations. Moreover, even 
if 
${\cal P}(\alpha)$ shows support for the baseline 
model $\alpha=1$, which lies within the 90\% CL of 
the distributions, EM observations seem to 
consistently favour larger values of the 3-body 
amplitude, reflecting stronger repulsive NNN 
interactions. As observed in \cite{Maselli:2020uol}, 
the distribution  of $\alpha$ inferred by GW data alone 
is unconstrained,
 with the posterior rallying against the lower prior at 
$\alpha=0.7$, while the multi-messenger analysis is 
dominated by the pulsar measurements, and in particular 
by J0740+6620, leading to values of $\alpha\gg 1$. 

Constraints on $\alpha$, i.e on the microscopic 
Hamiltonian \eqref{math:hamil}, 
can be translated into bounds on the stellar macroscopic 
observables. The right column of Fig.~\eqref{fig:3BC_real} 
shows, for example, the maximum mass density 
distributions predicted by the values of $\alpha$  
inferred for each dataset. 
All the observations lead to median values 
of $M_\tn{max}\gtrsim 2.2M_\odot$, with the multi-messenger 
analysis yielding a probability distribution with large 
support for $M_\tn{max}\sim 2.5M_\odot$. 

\begin{figure}[htbp!]
    \centering
    \includegraphics[scale = 0.45]{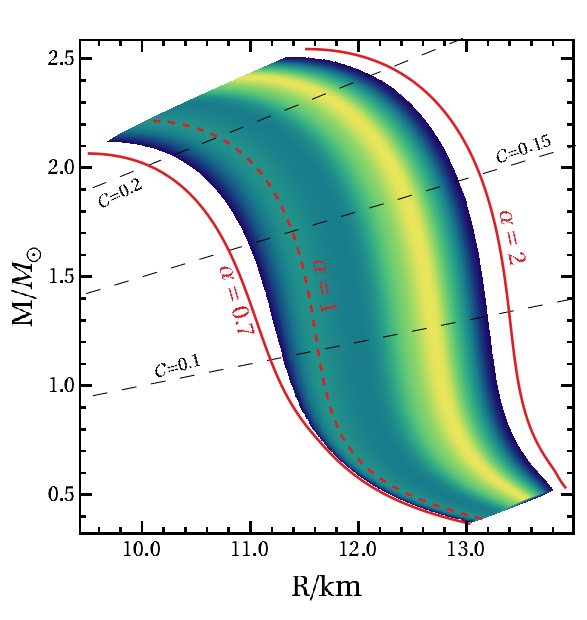}
    \caption{Mass-radius profile density corresponding 
    to the 90\% confidence interval of $\alpha$ 
    inferred for the GW-EM multi-messenger analysis. 
    Dark (light) regions correspond to stellar profiles 
    with small (large) probability. As for 
    Fig.~\ref{fig:eos_profiles} red curves identify 
    configurations with specific values of the 
    three-body strength, while dashed black lines correspond 
    to configurations with constant compactness.}
    \label{fig:MRCL}
\end{figure}

In Fig.~\ref{fig:MRCL} we also show the $M$-$R$ 
density distribution corresponding to the 
90\% CL of $\alpha$
for the multi-messenger case.  
Light (dark) colors identify stellar profiles 
with high (low) probability. Pulsar observations 
drive the profiles far from the $\alpha=1$ baseline, 
i.e. towards stiffer NS configurations, with an 
expected radius $R\gtrsim 12$ km for a prototype NS 
with $M=1.4M_\odot$. 

So far our analysis shows that, although the 
constraining power of current measurements is 
still limited, astrophysical data are already 
sensitive to nucleon dynamics. We will therefore 
explore the insights that can be inferred on 
three-body nuclear forces exploiting future GW 
observations of binary inspirals.

As discussed in Sec.~\ref{sec:sim} we have simulated two 
catalogues of $30$ binary NS mergers, observed either by 2G 
network or by ET, assuming two different 
values of the three-nucleon strength. Source parameters, 
i.e.~masses and tidal deformabilities, are first recovered 
with Bilby, and then analyzed by our Bayesian pipeline 
which samples the posterior distribution of $\alpha$. 

\begin{figure*}[htbp!]
    \centering
    \includegraphics[scale=0.33]{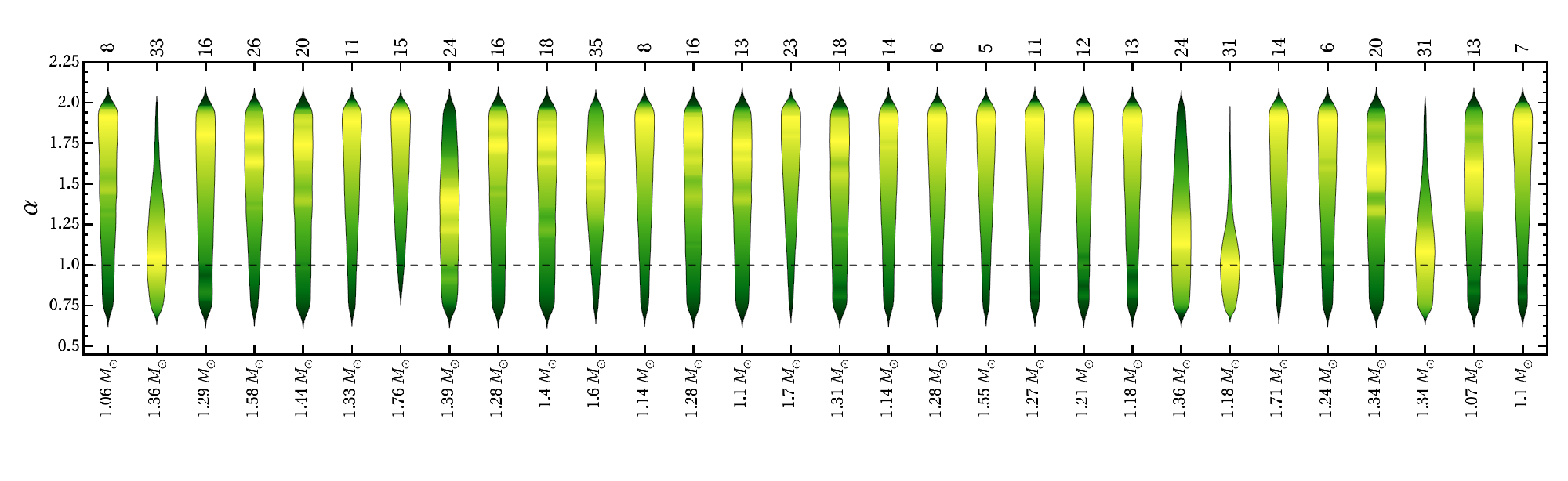}
    \caption{Posterior densities ${\cal P}(\alpha)$ inferred  
    from simulated GW data, assuming 
    $\alpha=1$ (dashed horizontal line). Yellow (green) 
    colors identify region with high (low) probability. 
    Signals are observed by a network HLV of three 
    advanced detectors, with a combined SNR given  
    in the top axis of the plot. Labels in the bottom 
    axis provide the values of the binary chirp masses.}
    \label{fig:HLV_violin}
\end{figure*}

Figure~\ref{fig:HLV_violin} shows the posterior densities 
${\cal P}(\alpha)$ of each event, for injected NSs with 
$\alpha=1$, detected by the HLV network. The 
ability of 2G detectors to discriminate the actual 
value of the three-body strength substantially depends on both 
the SNR and on the component masses of the binary. We find 
that observations with SNR smaller than $\sim 25$ lead 
$\alpha$ to be almost unconstrained, with the true value always 
lying outside the 90\% confidence interval of the distribution. 
However, even for strong signals, accurate measurements 
only occur for low-mass systems with a chirp mass 
${\cal M}\lesssim 1.4M_\odot$. This is particular evident 
for the event with the largest SNR ($\sim35$) in our set.  
Such binary features two heavy NSs with a chirp mass 
${\cal M}\simeq1.6M_\odot$, and provides loose  
bounds on $\alpha$. Moreover, 
Fig.~\ref{fig:HLV_violin} shows that, with the exception 
of four events with SNR$>30$ and ${\cal M}< 1.4M_\odot$, 
the remaining posteriors always prefer large values of the 
three-nucleon strength, at the edge of the upper prior boundary. 
This particular behavior reflects a systematic bias we find 
in the posteriors of $\tilde{\Lambda}$ inferred by GW 
observations for binaries with heavy components, which tend to 
favour large values of the tidal parameter. Its effect on 
the marginal distribution of $\alpha$ becomes even more 
pronounced in the high mass scenario where  the tidal 
deformability becomes less sensitive to variations of $\alpha$.  
We believe such bias may be induced by our choice of priors on the tidal 
parameters, which has strong support against the BBH 
hypothesis $\tilde{\Lambda}=0$, and reflects the 
physical assumption that compact objects with 
$m_{1,2}\lesssim3M_\odot$ are neutron stars.
Moreover, the stack of multiple GW signals only partially 
alleviate the bias in favour of large three-body strength. 
We have indeed combined different observations with SNR larger 
than 20, finding a mild improvement of the posterior 
support towards the true value of $\alpha$. The 
results discussed so far hold qualitatively also when 
we consider binary NSs simulated with $\alpha=1.3$.

\begin{figure*}[htbp!]
    \centering
    \includegraphics[scale = 0.33]{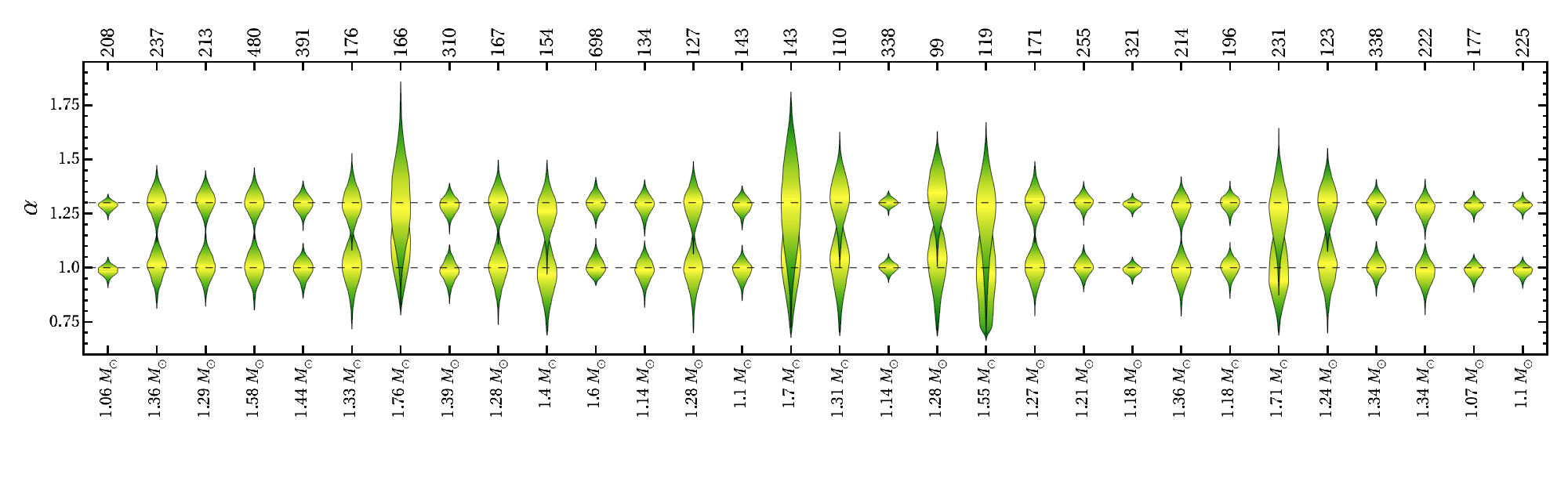}
    \caption{Same as Fig.~\ref{fig:HLV_violin} but 
    assuming that binary NS are observed by 
    the Einstein Telescope. We show results 
    for signals simulated with both $\alpha=1$ and $\alpha=1.3$. 
    Injected values of the three 
    body amplitude are identified by the horizontal 
    dashed lines.}
    \label{fig:ET_pdf}
\end{figure*}

\begin{figure}[htbp!]
    \centering
    \includegraphics[scale = 0.6]{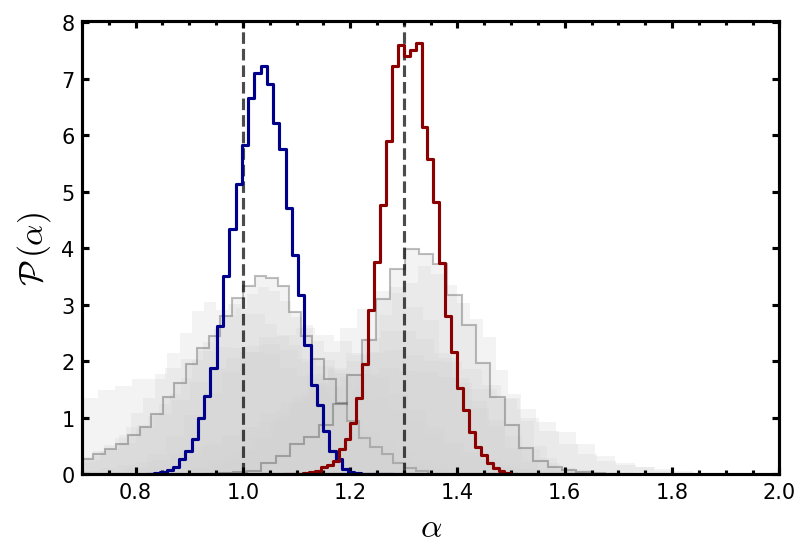}
    \caption{Probability distribution ${\cal P}(\alpha)$ 
    obtained by stacking six events of our dataset as 
    measured by the Einstein Telescope. 
    Empty histograms refer to the full 
    stacked posteriors for signals injected with 
    $\alpha=1$ and $\alpha=1.3$. Empty shaded 
    histograms on the background correspond 
    to the individual posteriors. The vertical 
    dashed lines identify the injected values 
    of $\alpha$.}
    \label{fig:ET_pdf_stack}
\end{figure}

This picture changes dramatically when signals are observed 
by the Einstein Telescope. Figure~\ref{fig:ET_pdf} shows 
indeed the distributions of the three-nucleon strength 
inferred by the 3G detector, for both families of events 
simulated with $\alpha=1$ and $\alpha=1.3$. The 
exquisite sensitivity of ET allows to gauge away 
the bias arising from the 2G network. All the 
posteriors peak around the injected 
values of $\alpha$, showing no support on the prior 
boundaries. In the best (worse) case scenario we find 
that $\alpha$ can 
be constrained with $\sim2\%$ ($\sim 30\%$) of 
accuracy at $68\%$ confidence level. 
Such accuracy allows to disentangle the two 
values of the three-body strength we consider. 
Even in the most pessimistic cases, where the 
inferred ${\cal P}(\alpha)$ are not narrow enough 
to identify a specific value of $\alpha$, stacking 
of few events would render the distributions 
clearly distinguishable. Figure \ref{fig:ET_pdf_stack} 
shows the posteriors obtained by combining six 
events of our catalogue\footnote{We choose the 
events number 7,15,16,18,19 and 25 of  Fig.~\ref{fig:ET_pdf}.}
leading to loose constraints on $\alpha$. 
The final posteriors for $\alpha=1$ and $\alpha=1.3$ 
are clearly separated, with a negligible overlap 
on the tails.

\begin{figure}[htp]
\centering
\includegraphics[scale=0.6]{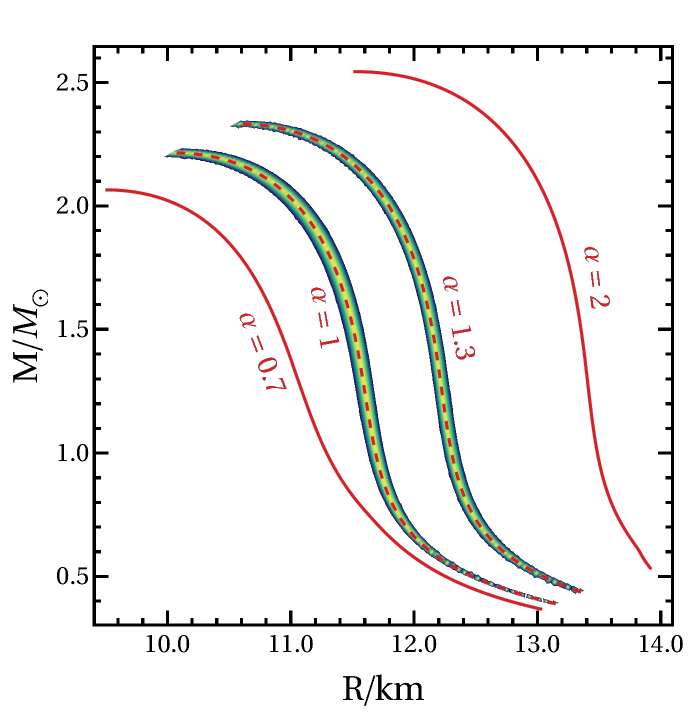}
\caption{Same as Fig.~\ref{fig:MRCL} but for simulated 
events observed by the Einstein Telescope. The values of 
$\alpha$ used to build the mass-radius profiles 
correspond correspond to event number 17 of our 
catalogue. We show results for both $\alpha=1$ 
and $\alpha=1.3$. Solid and dashed red curves identify 
the profiles corresponding to prior boundaries 
and to the injected values of $\alpha$, respectively.$\alpha$}
\label{fig:MRCLET}
\end{figure}

Such accuracy translates into very narrow constraints 
on the mass-radius (or equivalently mass-tidal deformability) 
diagram. As an example, we show in Fig.~\eqref{fig:MRCLET} 
the $M$-$R$ profile density computed from the values of $\alpha$ 
inferred from event number 17 of our dataset. 
A direct comparison with Fig.~\ref{fig:MRCL}, where 
a similar plot was made for data from current facilities, 
provides a clear hint on the possibility to use ET as 
a new laboratory to study the dynamics of 
nucleon interactions in the stellar cores.

\section{Conclusions}
\label{sec:summary}

We have investigated the sensitivity of NS observations to 
the strength of repulsive three-nucleon forces, which are 
known to be critical in determining the stiffness of the 
nuclear matter EOS at supranuclear densities. 
Our analysis is based on the AV18 + $\delta v$ + UIX$^*$ 
nuclear Hamiltonian and involves a single free parameter, 
to be constrained by data, determining the coupling constant 
appearing in the repulsive contribution to the UIX$^*$ potential. 

We have performed hierarchical bayesian inference employing the current available multimessenger datasets in order to constrain this parameter. We have then repeated the analysis with a set of simulated GW observations 
that could be performed by both current (LIGO/Virgo) and 
future (Einstein Telescope) interferometers at design sensitivity. 
This analysis has the main purpose to explore the potential of 
near and next generation facilities into inferring crucial 
information about the microscopic dynamics of nuclear matter.  

The analysis with real data has been carried out employing some of the 
dataset used in a previous work ~\cite{Maselli:2020uol}.
Our results suggest that even if current facilities show 
a clear sensitivity to small variation of the NNN repulsive 
potential, they are not accurate enough to 
capture significant insights. 
This picture is cross-validated by the population analysis 
performed with mocked LIGO/Virgo data, with binaries generated 
with two different values of the three-body strength, $\alpha=1$ 
and $\alpha=1.3$. Only few, low-mass and high SNR events provide 
a meaningful constraint on $\alpha$, with 
posterior distributions correctly peaked around the injected values.
Moreover, even for the most constraining event, the inferred 
posteriors do not allow a clear disentanglement between the 
two values of $\alpha$ we considered. The picture improves only 
slightly with the stacking of multiple observations. 

These results exhibit a striking upgrade when we assume that the 
population of binaries is observed by the Einstein Telescope. 
In most of the cases, the large SNRs obtained 
by such events in combination with the 3G detector allow the posteriors for the injected values of $\alpha$ to be clearly separated, and only a single observation is needed to 
resolve them.

Moreover, in the few cases where posteriors overlap, stacking of 
$\sim 2-3$ observations would allow to unambiguously 
distinguish between $\alpha=1$ 
and $\alpha=1.3$. The same conclusion would apply assuming that 
binaries are detected by the proposed Cosmic Explorer \cite{Essick:2017wyl,LIGOScientific:2016wof}.  
The large SNRs expected in the 3G era also 
require a careful assessment of waveform systematics 
which could bias the parameter reconstruction \cite{Chatziioannou:2021tdi,Gamba:2020wgg,Narikawa:2019xng,Castro:2022mpw}.
However, our results strongly support the evidence 
that with the upcoming third generation detectors, our 
understanding of neutron star matter will make a great step 
forward into the direction of using NS observations to probe 
fundamental physics at the fermi scale.

Further applications of our approach can 
be pursued following multiple directions, and in particular 
considering how constraints on nucleon dynamics would improve 
by joint analyses of the inspiral and of the post-merger phase, 
exploiting for the latter either GW oscillation modes \cite{Volkel:2022utc,Tonetto:2021ovc,Wijngaarden:2022sah}, or 
electromagnetic counterparts emitted by the binary remnant \cite{Breschi:2021tbm}.

\acknowledgments
Numerical calculations have been made possible through 
a CINECA-INFN agreement, providing access to resources on 
MARCONI at CINECA. We acknowledge financial support provided 
under the European Union's H2020 ERC, Starting Grant agreement 
no.~DarkGRA--757480. We also acknowledge support under the MIUR 
PRIN and FARE programmes (GW-NEXT, CUP:~B84I20000100001), and 
from the Amaldi Research Center funded by the MIUR 
program ``Dipartimento di Eccellenza'' (CUP: B81I18001170001). The work of O.B. and A.S. is supported by INFN through grant TEONGRAV.

\appendix*
\section{Parametrisation of energy density}

In this Appendix, we report the explicit expression of the energy density of nuclear 
matter employed to carry out our analysis. This expression was originally derived from a fit to the EOSs of SNM and PNM 
obtained by Akmal {\it et al.}~\cite{APR} using the AV18 + $\delta v$ + UIX$^*$ nuclear Hamiltonian and 
the variational FHNC/SOC formalism. 

The energy density of nuclear matter at baryon density $\varrho$ and proton fraction $x_p$ is written 
according to  Eqs.~\eqref{energy_fit1} and \eqref{energy_fit2}
\begin{align}
\nonumber
\epsilon(\varrho,x_p)  & = \Big[ \frac{\hbar^2}{2m}+ f(\varrho,x_p) \Big] \tau_p \\
 \label{happ} 
 & +  \Big[ \frac{\hbar^2}{2m}   + f(\varrho,1 - x_p) \Big] \tau_n \\
 \nonumber
 & + g(\varrho, 1/2) \Big[ 1 - (1 - 2 x_p)^2\Big] \\
 \nonumber
    &         + g(\rho, 0) (1 - 2 x_p)^2    , 
\end{align}
with 
\begin{align}
\label{taup}
\tau_p & = \varrho x_p \ \frac{3}{5} (3\pi^2\varrho x_p)^{2/3} \ , \\
\label{taun}
\tau_n & = \varrho (1-x_p) \ \frac{3}{5} [(3\pi^2\varrho (1-x_p)]^{2/3} \ .
\end{align}
The explicit form of the functions $f(\varrho,x_p)$ and $g(\varrho,x_p)$ appearing in Eq.~\eqref{happ} are
\begin{align}
\label{kin}
f(\varrho,x_p) = \left( a_1 + x_pa_2\right)\varrho e^{- a_3 \varrho}
\end{align}
and
\begin{align}
\label{pot1}
g(\varrho,x_p) = \left\{ 
\begin{tabular}{ll}
$g_L(\varrho,x_p)$ & $\varrho \le \bar{\varrho}$ \\
$g_H(\varrho,x_p)$ & $\varrho \ge \bar{\varrho}$
\end{tabular}
\right. \ , 
\end{align}
where
\begin{align}
\nonumber
g_L(\varrho, 1/2) & =  -\varrho^2\Big[ a_4+a_5\varrho + a_6\varrho^2 
  + (a_{7}+a_{8} \varrho) e^{-a_9^2\varrho^2}\Big] \ , \\
\label{pot2}
g_L(\rho, 0) & =  -\varrho^2\Big( a_{10} \varrho^{-1} + a_{11}+ a_{12} \varrho \Big)
       \ , \\
g_H(\varrho,1/2) & =  g_L(\varrho, 1/2) - \varrho^2 a_{13}(\varrho-a_{14}) e^{a_{15}(\rho-a_{14})}, \nonumber \\
g_H(\rho,0) & =  g_L(\rho, 0)
 - \varrho^2 a_{16}(\varrho-a_{17}) {\rm e}^{a_{18}(\varrho-a_{17})} \nonumber \ .
\end{align}
The density $\bar{\varrho} \lesssim 2 \varrho_0$ corresponds to the onset of the high-density 
phase\textemdash featuring spin-isospin density waves associated with neutral pion condensation\textemdash  
predicted by the study of Ref.~\cite{APR}.

The values of the parameters appearing in the above equations are given in Table~\ref{tab:par}
\begin{table*}[!h]

\caption
{Values of the parameters appearing in the definition of the energy density of  nuclear matter
of Eqs.~\eqref{happ}-~\eqref{pot2}, expressed in ${\rm MeV \ fm}^{-3}$.\label{tab:par}}
\vspace*{.10in}
\begin{tabular}{cccccc}
\hline\hline
  $a_1$ & $a_2$ & $a_3$ & $a_4$ & $a_5$ & $a_6$  \\
   $[{\rm MeV}\ {\rm fm}^5]$ & $[{\rm MeV}\ {\rm fm}^5]$ & $[{\rm fm}^3]$ & $[{\rm MeV}\ {\rm fm}^3]$ & $[{\rm MeV}\ {\rm fm}^6]$ & $[{\rm MeV}\ {\rm fm}^9]$ \\
   \noalign{\smallskip} \hline
 89.8 & -59.0 & 0.457 & 337.2 & -382. & -19.1 \\
\noalign{\smallskip\smallskip}
\hline \hline
  $a_7$ & $a_8$ & $a_9$ & $a_{10}$ & $a_{11}$ & $a_{12}$  \\
   $[{\rm MeV}\ {\rm fm}^3]$ & $[{\rm MeV}\ {\rm fm}^6]$ & $[{\rm fm}^3]$ & $[{\rm MeV}]$ & $[{\rm MeV}\ {\rm fm}^3]$ & $[{\rm MeV}\ {\rm fm}^6]$ \\
   \noalign{\smallskip} \hline
 69.0 & -33.0 & 6.4 & 0.35 & 214.6 & -384.0 \\
\noalign{\smallskip\smallskip}
\hline \hline
  $a_{13}$ & $a_{14}$ & $a_{15}$ & $a_{16}$ & $a_{17}$ & $a_{18}$  \\
   $[{\rm MeV}\ {\rm fm}^6]$ & $[{\rm fm}^{-3}]$ & $[{\rm MeV}\ {\rm fm}^6]$ & $[{\rm MeV}\ {\rm fm}^6]$ & $[{\rm MeV}]$ & $[{\rm fm}^3]$ \\
   \noalign{\smallskip} \hline
 175.0 & 0.32 & -1.45 & 287.0 & 0.195 & -1.54 \\
\hline \hline
\end{tabular}   
\end{table*}

\bibliographystyle{utphys}
\bibliography{Ref}

\providecommand{\href}[2]{#2}\begingroup\raggedright\begin{thebibliography}{10}

\bibitem{Cromartie:2019kug}
H.~T. Cromartie {\em et~al.}, ``{Relativistic Shapiro delay measurements of an
  extremely massive millisecond pulsar},''
  \href{http://arxiv.org/abs/1904.06759}{{\ttfamily arXiv:1904.06759}}.

\bibitem{Fonseca:2021wxt}
E.~Fonseca {\em et~al.}, ``{Refined Mass and Geometric Measurements of the
  High-mass PSR J0740+6620},''
  \href{http://dx.doi.org/10.3847/2041-8213/ac03b8}{{\em Astrophys. J. Lett.}
  {\bfseries 915} no.~1, (2021) L12},
  \href{http://arxiv.org/abs/2104.00880}{{\ttfamily arXiv:2104.00880
  [astro-ph.HE]}}.

\bibitem{Riley:2019yda}
T.~E. Riley {\em et~al.}, ``{A $NICER$ View of PSR J0030+0451: Millisecond
  Pulsar Parameter Estimation},''
  \href{http://dx.doi.org/10.3847/2041-8213/ab481c}{{\em Astrophys. J. Lett.}
  {\bfseries 887} no.~1, (2019) L21},
  \href{http://arxiv.org/abs/1912.05702}{{\ttfamily arXiv:1912.05702
  [astro-ph.HE]}}.

\bibitem{Miller:2019cac}
M.~C. Miller {\em et~al.}, ``{PSR J0030+0451 Mass and Radius from $NICER$ Data
  and Implications for the Properties of Neutron Star Matter},''
  \href{http://dx.doi.org/10.3847/2041-8213/ab50c5}{{\em Astrophys. J. Lett.}
  {\bfseries 887} no.~1, (2019) L24},
  \href{http://arxiv.org/abs/1912.05705}{{\ttfamily arXiv:1912.05705
  [astro-ph.HE]}}.

\bibitem{Riley:2021pdl}
T.~E. Riley {\em et~al.}, ``{A NICER View of the Massive Pulsar PSR J0740+6620
  Informed by Radio Timing and XMM-Newton Spectroscopy},''
  \href{http://dx.doi.org/10.3847/2041-8213/ac0a81}{{\em Astrophys. J. Lett.}
  {\bfseries 918} no.~2, (2021) L27},
  \href{http://arxiv.org/abs/2105.06980}{{\ttfamily arXiv:2105.06980
  [astro-ph.HE]}}.

\bibitem{Miller:2021qha}
M.~C. Miller {\em et~al.}, ``{The Radius of PSR J0740+6620 from NICER and
  XMM-Newton Data},'' \href{http://dx.doi.org/10.3847/2041-8213/ac089b}{{\em
  Astrophys. J. Lett.} {\bfseries 918} no.~2, (2021) L28},
  \href{http://arxiv.org/abs/2105.06979}{{\ttfamily arXiv:2105.06979
  [astro-ph.HE]}}.

\bibitem{LIGOScientific:2018hze}
{\bfseries LIGO Scientific, Virgo} Collaboration, B.~P. Abbott {\em et~al.},
  ``{Properties of the binary neutron star merger GW170817},''
  \href{http://dx.doi.org/10.1103/PhysRevX.9.011001}{{\em Phys. Rev. X}
  {\bfseries 9} no.~1, (2019) 011001},
  \href{http://arxiv.org/abs/1805.11579}{{\ttfamily arXiv:1805.11579 [gr-qc]}}.

\bibitem{LIGOScientific:2017vwq}
{\bfseries LIGO Scientific, Virgo} Collaboration, B.~P. Abbott {\em et~al.},
  ``{GW170817: Observation of Gravitational Waves from a Binary Neutron Star
  Inspiral},'' \href{http://dx.doi.org/10.1103/PhysRevLett.119.161101}{{\em
  Phys. Rev. Lett.} {\bfseries 119} no.~16, (2017) 161101},
  \href{http://arxiv.org/abs/1710.05832}{{\ttfamily arXiv:1710.05832 [gr-qc]}}.

\bibitem{LIGOScientific:2020aai}
{\bfseries LIGO Scientific, Virgo} Collaboration, B.~P. Abbott {\em et~al.},
  ``{GW190425: Observation of a Compact Binary Coalescence with Total Mass
  $\sim 3.4 M_{\odot}$},''
  \href{http://dx.doi.org/10.3847/2041-8213/ab75f5}{{\em Astrophys. J. Lett.}
  {\bfseries 892} no.~1, (2020) L3},
  \href{http://arxiv.org/abs/2001.01761}{{\ttfamily arXiv:2001.01761
  [astro-ph.HE]}}.

\bibitem{Hinderer:2007mb}
T.~Hinderer, ``{Tidal Love numbers of neutron stars},''
  \href{http://dx.doi.org/10.1086/533487}{{\em Astrophys. J.} {\bfseries 677}
  (2008) 1216--1220}, \href{http://arxiv.org/abs/0711.2420}{{\ttfamily
  arXiv:0711.2420 [astro-ph]}}.

\bibitem{Damour:2009vw}
T.~Damour and A.~Nagar, ``{Relativistic tidal properties of neutron stars},''
  \href{http://dx.doi.org/10.1103/PhysRevD.80.084035}{{\em Phys. Rev. D}
  {\bfseries 80} (2009) 084035},
  \href{http://arxiv.org/abs/0906.0096}{{\ttfamily arXiv:0906.0096 [gr-qc]}}.

\bibitem{Binnington:2009bb}
T.~Binnington and E.~Poisson, ``{Relativistic theory of tidal Love numbers},''
  \href{http://dx.doi.org/10.1103/PhysRevD.80.084018}{{\em Phys. Rev. D}
  {\bfseries 80} (2009) 084018},
  \href{http://arxiv.org/abs/0906.1366}{{\ttfamily arXiv:0906.1366 [gr-qc]}}.

\bibitem{Flanagan:2007ix}
E.~E. Flanagan and T.~Hinderer, ``{Constraining neutron star tidal Love numbers
  with gravitational wave detectors},''
  \href{http://dx.doi.org/10.1103/PhysRevD.77.021502}{{\em Phys. Rev. D}
  {\bfseries 77} (2008) 021502},
  \href{http://arxiv.org/abs/0709.1915}{{\ttfamily arXiv:0709.1915
  [astro-ph]}}.

\bibitem{Vines:2010ca}
J.~E. Vines and E.~E. Flanagan, ``{Post-1-Newtonian quadrupole tidal
  interactions in binary systems},''
  \href{http://dx.doi.org/10.1103/PhysRevD.88.024046}{{\em Phys. Rev. D}
  {\bfseries 88} (2013) 024046},
  \href{http://arxiv.org/abs/1009.4919}{{\ttfamily arXiv:1009.4919 [gr-qc]}}.

\bibitem{Vines:2011ud}
J.~Vines, E.~E. Flanagan, and T.~Hinderer, ``{Post-1-Newtonian tidal effects in
  the gravitational waveform from binary inspirals},''
  \href{http://dx.doi.org/10.1103/PhysRevD.83.084051}{{\em Phys. Rev. D}
  {\bfseries 83} (2011) 084051},
  \href{http://arxiv.org/abs/1101.1673}{{\ttfamily arXiv:1101.1673 [gr-qc]}}.

\bibitem{2008PPN....39..286C}
G.~{Col{\`o}}, ``{The compression modes in atomic nuclei and their relevance
  for the nuclear equation of state},''
  \href{http://dx.doi.org/10.1007/s11496-008-2005-2}{{\em Physics of Particles
  and Nuclei} {\bfseries 39} no.~2, (Mar., 2008) 286--305}.

\bibitem{Li:2013ola}
B.-A. Li and X.~Han, ``{Constraining the neutron-proton effective mass
  splitting using empirical constraints on the density dependence of nuclear
  symmetry energy around normal density},''
  \href{http://dx.doi.org/10.1016/j.physletb.2013.10.006}{{\em Phys. Lett. B}
  {\bfseries 727} (2013) 276--281},
  \href{http://arxiv.org/abs/1304.3368}{{\ttfamily arXiv:1304.3368 [nucl-th]}}.

\bibitem{Russotto:2016ucm}
P.~Russotto {\em et~al.}, ``{Results of the ASY-EOS experiment at GSI: The
  symmetry energy at suprasaturation density},''
  \href{http://dx.doi.org/10.1103/PhysRevC.94.034608}{{\em Phys. Rev. C}
  {\bfseries 94} no.~3, (2016) 034608},
  \href{http://arxiv.org/abs/1608.04332}{{\ttfamily arXiv:1608.04332
  [nucl-ex]}}.

\bibitem{Tsang:2008fd}
M.~B. Tsang, Y.~Zhang, P.~Danielewicz, M.~Famiano, Z.~Li, W.~G. Lynch, and
  A.~W. Steiner, ``{Constraints on the density dependence of the symmetry
  energy},'' \href{http://dx.doi.org/10.1103/PhysRevLett.102.122701}{{\em Phys.
  Rev. Lett.} {\bfseries 102} (2009) 122701},
  \href{http://arxiv.org/abs/0811.3107}{{\ttfamily arXiv:0811.3107 [nucl-ex]}}.

\bibitem{Danielewicz:2002pu}
P.~Danielewicz, R.~Lacey, and W.~G. Lynch, ``{Determination of the equation of
  state of dense matter},''
  \href{http://dx.doi.org/10.1126/science.1078070}{{\em Science} {\bfseries
  298} (2002) 1592--1596},
  \href{http://arxiv.org/abs/nucl-th/0208016}{{\ttfamily
  arXiv:nucl-th/0208016}}.

\bibitem{Brown:2013mga}
B.~A. Brown, ``{Constraints on the Skyrme Equations of State from Properties of
  Doubly Magic Nuclei},''
  \href{http://dx.doi.org/10.1103/PhysRevLett.111.232502}{{\em Phys. Rev.
  Lett.} {\bfseries 111} no.~23, (2013) 232502},
  \href{http://arxiv.org/abs/1308.3664}{{\ttfamily arXiv:1308.3664 [nucl-th]}}.

\bibitem{Zhang:2013wna}
Z.~Zhang and L.-W. Chen, ``{Constraining the symmetry energy at subsaturation
  densities using isotope binding energy difference and neutron skin
  thickness},'' \href{http://dx.doi.org/10.1016/j.physletb.2013.08.002}{{\em
  Phys. Lett. B} {\bfseries 726} (2013) 234--238},
  \href{http://arxiv.org/abs/1302.5327}{{\ttfamily arXiv:1302.5327 [nucl-th]}}.

\bibitem{PREX:2021umo}
{\bfseries PREX} Collaboration, D.~Adhikari {\em et~al.}, ``{Accurate
  Determination of the Neutron Skin Thickness of $^{208}$Pb through
  Parity-Violation in Electron Scattering},''
  \href{http://dx.doi.org/10.1103/PhysRevLett.126.172502}{{\em Phys. Rev.
  Lett.} {\bfseries 126} no.~17, (2021) 172502},
  \href{http://arxiv.org/abs/2102.10767}{{\ttfamily arXiv:2102.10767
  [nucl-ex]}}.

\bibitem{Annala:2017llu}
E.~Annala, T.~Gorda, A.~Kurkela, and A.~Vuorinen, ``{Gravitational-wave
  constraints on the neutron-star-matter Equation of State},''
  \href{http://dx.doi.org/10.1103/PhysRevLett.120.172703}{{\em Phys. Rev.
  Lett.} {\bfseries 120} no.~17, (2018) 172703},
  \href{http://arxiv.org/abs/1711.02644}{{\ttfamily arXiv:1711.02644
  [astro-ph.HE]}}.

\bibitem{Margalit:2017dij}
B.~Margalit and B.~D. Metzger, ``{Constraining the Maximum Mass of Neutron
  Stars From Multi-Messenger Observations of GW170817},''
  \href{http://dx.doi.org/10.3847/2041-8213/aa991c}{{\em Astrophys. J. Lett.}
  {\bfseries 850} no.~2, (2017) L19},
  \href{http://arxiv.org/abs/1710.05938}{{\ttfamily arXiv:1710.05938
  [astro-ph.HE]}}.

\bibitem{Radice:2017lry}
D.~Radice, A.~Perego, F.~Zappa, and S.~Bernuzzi, ``{GW170817: Joint Constraint
  on the Neutron Star Equation of State from Multimessenger Observations},''
  \href{http://dx.doi.org/10.3847/2041-8213/aaa402}{{\em Astrophys. J. Lett.}
  {\bfseries 852} no.~2, (2018) L29},
  \href{http://arxiv.org/abs/1711.03647}{{\ttfamily arXiv:1711.03647
  [astro-ph.HE]}}.

\bibitem{Bauswein:2017vtn}
A.~Bauswein, O.~Just, H.-T. Janka, and N.~Stergioulas, ``{Neutron-star radius
  constraints from GW170817 and future detections},''
  \href{http://dx.doi.org/10.3847/2041-8213/aa9994}{{\em Astrophys. J. Lett.}
  {\bfseries 850} no.~2, (2017) L34},
  \href{http://arxiv.org/abs/1710.06843}{{\ttfamily arXiv:1710.06843
  [astro-ph.HE]}}.

\bibitem{Lim:2018bkq}
Y.~Lim and J.~W. Holt, ``{Neutron star tidal deformabilities constrained by
  nuclear theory and experiment},''
  \href{http://dx.doi.org/10.1103/PhysRevLett.121.062701}{{\em Phys. Rev.
  Lett.} {\bfseries 121} no.~6, (2018) 062701},
  \href{http://arxiv.org/abs/1803.02803}{{\ttfamily arXiv:1803.02803
  [nucl-th]}}.

\bibitem{Lim:2020zvx}
Y.~Lim, A.~Bhattacharya, J.~W. Holt, and D.~Pati, ``{Radius and equation of
  state constraints from massive neutron stars and GW190814},''
  \href{http://dx.doi.org/10.1103/PhysRevC.104.L032802}{{\em Phys. Rev. C}
  {\bfseries 104} no.~3, (2021) L032802},
  \href{http://arxiv.org/abs/2007.06526}{{\ttfamily arXiv:2007.06526
  [nucl-th]}}.

\bibitem{Most:2018hfd}
E.~R. Most, L.~R. Weih, L.~Rezzolla, and J.~Schaffner-Bielich, ``{New
  constraints on radii and tidal deformabilities of neutron stars from
  GW170817},'' \href{http://dx.doi.org/10.1103/PhysRevLett.120.261103}{{\em
  Phys. Rev. Lett.} {\bfseries 120} no.~26, (2018) 261103},
  \href{http://arxiv.org/abs/1803.00549}{{\ttfamily arXiv:1803.00549 [gr-qc]}}.

\bibitem{De:2018uhw}
S.~De, D.~Finstad, J.~M. Lattimer, D.~A. Brown, E.~Berger, and C.~M. Biwer,
  ``{Tidal Deformabilities and Radii of Neutron Stars from the Observation of
  GW170817},'' \href{http://dx.doi.org/10.1103/PhysRevLett.121.091102}{{\em
  Phys. Rev. Lett.} {\bfseries 121} no.~9, (2018) 091102},
  \href{http://arxiv.org/abs/1804.08583}{{\ttfamily arXiv:1804.08583
  [astro-ph.HE]}}. [Erratum: Phys.Rev.Lett. 121, 259902 (2018)].

\bibitem{Annala:2019puf}
E.~Annala, T.~Gorda, A.~Kurkela, J.~N\"attil\"a, and A.~Vuorinen, ``{Evidence
  for quark-matter cores in massive neutron stars},''
  \href{http://dx.doi.org/10.1038/s41567-020-0914-9}{{\em Nature Phys.}
  {\bfseries 16} no.~9, (2020) 907--910},
  \href{http://arxiv.org/abs/1903.09121}{{\ttfamily arXiv:1903.09121
  [astro-ph.HE]}}.

\bibitem{Raaijmakers:2019dks}
G.~Raaijmakers {\em et~al.}, ``{Constraining the dense matter equation of state
  with joint analysis of NICER and LIGO/Virgo measurements},''
  \href{http://dx.doi.org/10.3847/2041-8213/ab822f}{{\em Astrophys. J. Lett.}
  {\bfseries 893} no.~1, (2020) L21},
  \href{http://arxiv.org/abs/1912.11031}{{\ttfamily arXiv:1912.11031
  [astro-ph.HE]}}.

\bibitem{Miller:2019nzo}
M.~C. Miller, C.~Chirenti, and F.~K. Lamb, ``{Constraining the equation of
  state of high-density cold matter using nuclear and astronomical
  measurements},'' \href{http://arxiv.org/abs/1904.08907}{{\ttfamily
  arXiv:1904.08907 [astro-ph.HE]}}.

\bibitem{Kumar:2019xgp}
B.~Kumar and P.~Landry, ``{Inferring neutron star properties from GW170817 with
  universal relations},''
  \href{http://dx.doi.org/10.1103/PhysRevD.99.123026}{{\em Phys. Rev. D}
  {\bfseries 99} no.~12, (2019) 123026},
  \href{http://arxiv.org/abs/1902.04557}{{\ttfamily arXiv:1902.04557 [gr-qc]}}.

\bibitem{Fasano:2019zwm}
M.~Fasano, T.~Abdelsalhin, A.~Maselli, and V.~Ferrari, ``{Constraining the
  Neutron Star Equation of State Using Multiband Independent Measurements of
  Radii and Tidal Deformabilities},''
  \href{http://dx.doi.org/10.1103/PhysRevLett.123.141101}{{\em Phys. Rev.
  Lett.} {\bfseries 123} no.~14, (2019) 141101},
  \href{http://arxiv.org/abs/1902.05078}{{\ttfamily arXiv:1902.05078
  [astro-ph.HE]}}.

\bibitem{Landry:2020vaw}
P.~Landry, R.~Essick, and K.~Chatziioannou, ``{Nonparametric constraints on
  neutron star matter with existing and upcoming gravitational wave and pulsar
  observations},'' \href{http://dx.doi.org/10.1103/PhysRevD.101.123007}{{\em
  Phys. Rev. D} {\bfseries 101} no.~12, (2020) 123007},
  \href{http://arxiv.org/abs/2003.04880}{{\ttfamily arXiv:2003.04880
  [astro-ph.HE]}}.

\bibitem{Guven:2020dok}
H.~G\"uven, K.~Bozkurt, E.~Khan, and J.~Margueron, ``{Multimessenger and
  multiphysics Bayesian inference for the GW170817 binary neutron star
  merger},'' \href{http://dx.doi.org/10.1103/PhysRevC.102.015805}{{\em Phys.
  Rev. C} {\bfseries 102} no.~1, (2020) 015805},
  \href{http://arxiv.org/abs/2001.10259}{{\ttfamily arXiv:2001.10259
  [nucl-th]}}.

\bibitem{Traversi:2020aaa}
S.~Traversi, P.~Char, and G.~Pagliara, ``{Bayesian Inference of Dense Matter
  Equation of State within Relativistic Mean Field Models using Astrophysical
  Measurements},'' \href{http://dx.doi.org/10.3847/1538-4357/ab99c1}{{\em
  Astrophys. J.} {\bfseries 897} (2020) 165},
  \href{http://arxiv.org/abs/2002.08951}{{\ttfamily arXiv:2002.08951
  [astro-ph.HE]}}.

\bibitem{Raaijmakers:2021uju}
G.~Raaijmakers, S.~K. Greif, K.~Hebeler, T.~Hinderer, S.~Nissanke, A.~Schwenk,
  T.~E. Riley, A.~L. Watts, J.~M. Lattimer, and W.~C.~G. Ho, ``{Constraints on
  the Dense Matter Equation of State and Neutron Star Properties from
  NICER\textquoteright{}s Mass\textendash{}Radius Estimate of PSR J0740+6620
  and Multimessenger Observations},''
  \href{http://dx.doi.org/10.3847/2041-8213/ac089a}{{\em Astrophys. J. Lett.}
  {\bfseries 918} no.~2, (2021) L29},
  \href{http://arxiv.org/abs/2105.06981}{{\ttfamily arXiv:2105.06981
  [astro-ph.HE]}}.

\bibitem{Zimmerman:2020eho}
J.~Zimmerman, Z.~Carson, K.~Schumacher, A.~W. Steiner, and K.~Yagi,
  ``{Measuring Nuclear Matter Parameters with NICER and LIGO/Virgo},''
  \href{http://arxiv.org/abs/2002.03210}{{\ttfamily arXiv:2002.03210
  [astro-ph.HE]}}.

\bibitem{Silva:2020acr}
H.~O. Silva, A.~M. Holgado, A.~C\'ardenas-Avenda\~no, and N.~Yunes,
  ``{Astrophysical and theoretical physics implications from multimessenger
  neutron star observations},''
  \href{http://dx.doi.org/10.1103/PhysRevLett.126.181101}{{\em Phys. Rev.
  Lett.} {\bfseries 126} no.~18, (2021) 181101},
  \href{http://arxiv.org/abs/2004.01253}{{\ttfamily arXiv:2004.01253 [gr-qc]}}.

\bibitem{Sabatucci:2020xwt}
A.~Sabatucci and O.~Benhar, ``{Tidal Deformation of Neutron Stars from
  Microscopic Models of Nuclear Dynamics},''
  \href{http://dx.doi.org/10.1103/PhysRevC.101.045807}{{\em Phys. Rev. C}
  {\bfseries 101} no.~4, (2020) 045807},
  \href{http://arxiv.org/abs/2001.06294}{{\ttfamily arXiv:2001.06294
  [nucl-th]}}.

\bibitem{Blaschke:2020qqj}
D.~Blaschke, A.~Ayriyan, D.~E. Alvarez-Castillo, and H.~Grigorian, ``{Was
  GW170817 a Canonical Neutron Star Merger? Bayesian Analysis with a Third
  Family of Compact Stars},''
  \href{http://dx.doi.org/10.3390/universe6060081}{{\em Universe} {\bfseries 6}
  no.~6, (2020) 81}, \href{http://arxiv.org/abs/2005.02759}{{\ttfamily
  arXiv:2005.02759 [astro-ph.HE]}}.

\bibitem{Tang:2020koz}
S.-P. Tang, J.-L. Jiang, W.-H. Gao, Y.-Z. Fan, and D.-M. Wei, ``{Constraint on
  phase transition with the multimessenger data of neutron stars},''
  \href{http://dx.doi.org/10.1103/PhysRevD.103.063026}{{\em Phys. Rev. D}
  {\bfseries 103} no.~6, (2021) 063026},
  \href{http://arxiv.org/abs/2009.05719}{{\ttfamily arXiv:2009.05719
  [astro-ph.HE]}}.

\bibitem{Biswas:2020puz}
B.~Biswas, P.~Char, R.~Nandi, and S.~Bose, ``{Towards mitigation of apparent
  tension between nuclear physics and astrophysical observations by improved
  modeling of neutron star matter},''
  \href{http://dx.doi.org/10.1103/PhysRevD.103.103015}{{\em Phys. Rev. D}
  {\bfseries 103} no.~10, (2021) 103015},
  \href{http://arxiv.org/abs/2008.01582}{{\ttfamily arXiv:2008.01582
  [astro-ph.HE]}}.

\bibitem{Pacilio:2021jmq}
C.~Pacilio, A.~Maselli, M.~Fasano, and P.~Pani, ``{Ranking Love Numbers for the
  Neutron Star Equation of State: The Need for Third-Generation Detectors},''
  \href{http://dx.doi.org/10.1103/PhysRevLett.128.101101}{{\em Phys. Rev.
  Lett.} {\bfseries 128} no.~10, (2022) 101101},
  \href{http://arxiv.org/abs/2104.10035}{{\ttfamily arXiv:2104.10035 [gr-qc]}}.

\bibitem{Malik:2022jqc}
T.~Malik and C.~Provid\^encia, ``{Bayesian inference of signatures of hyperons
  inside neutron stars},'' \href{http://arxiv.org/abs/2205.15843}{{\ttfamily
  arXiv:2205.15843 [nucl-th]}}.

\bibitem{Altiparmak:2022bke}
S.~Altiparmak, C.~Ecker, and L.~Rezzolla, ``{On the Sound Speed in Neutron
  Stars},'' \href{http://arxiv.org/abs/2203.14974}{{\ttfamily arXiv:2203.14974
  [astro-ph.HE]}}.

\bibitem{Gupta:2022qgg}
P.~K. Gupta, A.~Puecher, P.~T.~H. Pang, J.~Janquart, G.~Koekoek, and C.~Broeck
  Van~Den, ``{Determining the equation of state of neutron stars with Einstein
  Telescope using tidal effects and r-mode excitations from a population of
  binary inspirals},'' \href{http://arxiv.org/abs/2205.01182}{{\ttfamily
  arXiv:2205.01182 [gr-qc]}}.

\bibitem{Baiotti:2019sew}
L.~Baiotti, ``{Gravitational waves from neutron star mergers and their relation
  to the nuclear equation of state},''
  \href{http://dx.doi.org/10.1016/j.ppnp.2019.103714}{{\em Prog. Part. Nucl.
  Phys.} {\bfseries 109} (2019) 103714},
  \href{http://arxiv.org/abs/1907.08534}{{\ttfamily arXiv:1907.08534
  [astro-ph.HE]}}.

\bibitem{Chatziioannou:2020pqz}
K.~Chatziioannou, ``{Neutron star tidal deformability and equation of state
  constraints},'' \href{http://dx.doi.org/10.1007/s10714-020-02754-3}{{\em Gen.
  Rel. Grav.} {\bfseries 52} no.~11, (2020) 109},
  \href{http://arxiv.org/abs/2006.03168}{{\ttfamily arXiv:2006.03168 [gr-qc]}}.

\bibitem{Maselli:2020uol}
A.~Maselli, A.~Sabatucci, and O.~Benhar, ``{Constraining three-nucleon forces
  with multimessenger data},''
  \href{http://dx.doi.org/10.1103/PhysRevC.103.065804}{{\em Phys. Rev. C}
  {\bfseries 103} no.~6, (2021) 065804},
  \href{http://arxiv.org/abs/2010.03581}{{\ttfamily arXiv:2010.03581
  [astro-ph.HE]}}.

\bibitem{APR}
{A. Akmal, V.R. Pandharipande, and D.G. Ravenhall}, ``Equation of state of
  nucleon matter and neutron star structure,'' {\em Phys. Rev. C} {\bfseries
  58} (1998) 1804.

\bibitem{UIX_1}
B.~S. Pudliner, V.~R. Pandharipande, J.~Carlson, and R.~B. Wiringa, ``{Quantum
  Monte Carlo calculations of A <= 6 nuclei},'' {\em Phys. Rev. Lett.}
  {\bfseries 74} (1995) 4396--4399.

\bibitem{UIX_2}
J.~Carlson, V.~R. Pandharipande, and R.~B. Wiringa {\em Nucl. Phys. A}
  {\bfseries 401} (1983) 59.

\bibitem{Punturo:2010zz}
M.~Punturo {\em et~al.}, ``{The Einstein Telescope: A third-generation
  gravitational wave observatory},''
  \href{http://dx.doi.org/10.1088/0264-9381/27/19/194002}{{\em Class. Quant.
  Grav.} {\bfseries 27} (2010) 194002}.

\bibitem{Hild:2010id}
S.~Hild {\em et~al.}, ``{Sensitivity Studies for Third-Generation Gravitational
  Wave Observatories},''
  \href{http://dx.doi.org/10.1088/0264-9381/28/9/094013}{{\em Class. Quant.
  Grav.} {\bfseries 28} (2011) 094013},
  \href{http://arxiv.org/abs/1012.0908}{{\ttfamily arXiv:1012.0908 [gr-qc]}}.

\bibitem{Maggiore:2019uih}
M.~Maggiore {\em et~al.}, ``{Science Case for the Einstein Telescope},''
  \href{http://dx.doi.org/10.1088/1475-7516/2020/03/050}{{\em JCAP} {\bfseries
  03} (2020) 050}, \href{http://arxiv.org/abs/1912.02622}{{\ttfamily
  arXiv:1912.02622 [astro-ph.CO]}}.

\bibitem{QMC}
J.~Carlson, S.~Gandolfi, F.~Pederiva, S.~C. Pieper, R.~Schiavilla, K.~E.
  Schmidt, and R.~B. Wiringa {\em Rev. Mod. Phys.} {\bfseries 87} (2015) 1067.

\bibitem{Benhar:IJMPE}
O.~Benhar {\em Int. J. Mod. Phys. E} {\bfseries 9} (2021) 2130009.

\bibitem{Essick2020}
R.~Essick, I.~Tews, P.~Landry, S.~Reddy, and D.~E. Holz, ``Direct astrophysical
  tests of chiral effective field theory at supranuclear densities,''
  \href{http://dx.doi.org/10.1103/PhysRevC.102.055803}{{\em Phys. Rev. C}
  {\bfseries 102} (Nov, 2020) 055803}.
  \url{https://link.aps.org/doi/10.1103/PhysRevC.102.055803}.

\bibitem{AV18}
R.~B. Wiringa, V.~G.~J. Stoks, and R.~Schiavilla, ``{An Accurate
  nucleon-nucleon potential with charge independence breaking},'' {\em Phys.
  Rev. C} {\bfseries 51} (1995) 38--51.

\bibitem{Akmal:1997}
A.~Akmal and V.~R. Pandharipande, ``Spin-isospin structure and pion
  condensation in nucleon matter,''
  \href{http://dx.doi.org/10.1103/PhysRevC.56.2261}{{\em Phys. Rev. C}
  {\bfseries 56} (Oct, 1997) 2261--2279}.
  \url{https://link.aps.org/doi/10.1103/PhysRevC.56.2261}.

\bibitem{T}
{Tolman, R. C.}, ``Static solutions of {Einstein's} field equations for spheres
  of fluid,'' {\em Phys. Rev.} {\bfseries 55} (1939) 364.

\bibitem{OV}
{Oppenheimer, J. R. and Volkoff, G. M.}, ``On massive neutron cores,'' {\em
  Phys. Rev.} {\bfseries 55} (1939) 374.

\bibitem{Sabatucci2020}
A.~Sabatucci and O.~Benhar, ``Tidal deformation of neutron stars from
  microscopic models of nuclear dynamics,''
  \href{http://dx.doi.org/10.1103/PhysRevC.101.045807}{{\em Phys. Rev. C}
  {\bfseries 101} (Apr, 2020) 045807}.
  \url{https://link.aps.org/doi/10.1103/PhysRevC.101.045807}.

\bibitem{Fujita}
J.~Fujita and H.~Miyazawa, ``{Pion Theory of Three-Body Forces},'' {\em Prog.
  Theor. Phys.} {\bfseries 17} (1957) 360--365.

\bibitem{boost}
J.~Forest, V.~R. Pandharipande, and J.~L. Friar {\em Phys. Rev. C} {\bfseries
  52} (1995) 568.

\bibitem{bob_vijay_rmp}
V.~R. Pandharipande and R.~B. Wiringa {\em Rev. Mod. Phys.} {\bfseries 51}
  (1979) 821.

\bibitem{BL:2017}
{ O. Benhar and A. Lovato}, ``Perturbation theory of nuclear matter with a
  microscopic effective interaction,'' {\em Phys. Rev. C} {\bfseries 96} (2017)
  054301.

\bibitem{Carlson:QMC}
J.~Carlson, S.~Gandolfi, F.~Pederiva, S.~C. Pieper, R.~Schiavilla, K.~E.
  Schmidt, and R.~B. Wiringa, ``Quantum monte carlo methods for nuclear
  physics,'' \href{http://dx.doi.org/10.1103/RevModPhys.87.1067}{{\em Rev. Mod.
  Phys.} {\bfseries 87} (Sep, 2015) 1067--1118}.
  \url{https://link.aps.org/doi/10.1103/RevModPhys.87.1067}.

\bibitem{2013PASP..125..306F}
D.~{Foreman-Mackey}, D.~W. {Hogg}, D.~{Lang}, and J.~{Goodman}, ``{emcee: The
  MCMC Hammer},'' \href{http://dx.doi.org/10.1086/670067}{{\em \pasp}
  {\bfseries 125} no.~925, (Mar., 2013) 306},
  \href{http://arxiv.org/abs/1202.3665}{{\ttfamily arXiv:1202.3665
  [astro-ph.IM]}}.

\bibitem{riley_thomas_e_2020_5506838}
T.~E. Riley, A.~L. Watts, S.~Bogdanov, P.~S. Ray, R.~M. Ludlam, S.~Guillot,
  Z.~Arzoumanian, C.~L. Baker, A.~V. Bilous, D.~Chakrabarty, K.~C. Gendreau,
  A.~K. Harding, W.~C.~G. Ho, J.~M. Lattimer, S.~M. Morsink, and T.~E.
  Strohmayer, ``{A NICER View of PSR J0030+0451: Nested Samples for Millisecond
  Pulsar Parameter Estimation},'' Mar., 2020.
\newblock \url{https://doi.org/10.5281/zenodo.5506838}.

\bibitem{riley_thomas_e_2021_4697625}
T.~E. Riley, A.~L. Watts, P.~S. Ray, S.~Bogdanov, S.~Guillot, S.~M. Morsink,
  A.~V. Bilous, Z.~Arzoumanian, D.~Choudhury, J.~S. Deneva, K.~C. Gendreau,
  A.~K. Harding, W.~C. Ho, J.~M. Lattimer, M.~Loewenstein, R.~M. Ludlam, C.~B.
  Markwardt, T.~Okajima, C.~Prescod-Weinstein, R.~A. Remillard, M.~T. Wolff,
  E.~Fonseca, H.~T. Cromartie, M.~Kerr, T.~T. Pennucci, A.~Parthasarathy,
  S.~Ransom, I.~Stairs, L.~Guillemot, and I.~Cognard, ``{A NICER View of the
  Massive Pulsar PSR J0740+6620 Informed by Radio Timing and XMM-Newton
  Spectroscopy: Nested Samples for Millisecond Pulsar Parameter Estimation},''
  Apr., 2021.
\newblock \url{https://doi.org/10.5281/zenodo.4697625}.

\bibitem{miller_m_c_2021_4670689}
M.~Miller, F.~K. Lamb, A.~J. Dittmann, S.~Bogdanov, Z.~Arzoumanian, K.~C.
  Gendreau, S.~Guillot, W.~C.~G. Ho, J.~M. Lattimer, S.~M. Morsink, P.~S. Ray,
  M.~T. Wolff, C.~L. Baker, T.~Cazeau, S.~Manthripragada, C.~B. Markwardt,
  T.~Okajima, S.~Pollard, I.~Cognard, H.~T. Cromartie, E.~Fonseca,
  L.~Guillemot, M.~Kerr, A.~Parthasarathy, T.~T. Pennucci, S.~Ransom,
  I.~Stairs, and M.~Loewenstein, ``{NICER PSR J0740+6620 Illinois-Maryland MCMC
  Samples},'' Apr., 2021.
\newblock \url{https://doi.org/10.5281/zenodo.4670689}.

\bibitem{raaijmakers_g_2021_4696232}
G.~Raaijmakers, S.~Greif, K.~Hebeler, T.~Hinderer, S.~Nissanke, A.~Schwenk,
  T.~Riley, J.~Lattimer, and W.~Ho, ``{Constraints on the dense matter equation
  of state and neutron star properties from NICER's mass- radius estimate of
  PSR J0740+6620 and multimessenger observations: posterior samples and scripts
  for generating plots},'' Apr., 2021.
\newblock \url{https://doi.org/10.5281/zenodo.4696232}.

\bibitem{Wysocki:2020myz}
D.~Wysocki, R.~O'Shaughnessy, L.~Wade, and J.~Lange, ``{Inferring the neutron
  star equation of state simultaneously with the population of merging neutron
  stars},'' \href{http://arxiv.org/abs/2001.01747}{{\ttfamily arXiv:2001.01747
  [gr-qc]}}.

\bibitem{Pitkin:2011yk}
M.~Pitkin, S.~Reid, S.~Rowan, and J.~Hough, ``{Gravitational Wave Detection by
  Interferometry (Ground and Space)},''
  \href{http://dx.doi.org/10.12942/lrr-2011-5}{{\em Living Rev. Rel.}
  {\bfseries 14} (2011) 5}, \href{http://arxiv.org/abs/1102.3355}{{\ttfamily
  arXiv:1102.3355 [astro-ph.IM]}}.

\bibitem{Wade:2014vqa}
L.~Wade, J.~D.~E. Creighton, E.~Ochsner, B.~D. Lackey, B.~F. Farr, T.~B.
  Littenberg, and V.~Raymond, ``{Systematic and statistical errors in a
  bayesian approach to the estimation of the neutron-star equation of state
  using advanced gravitational wave detectors},''
  \href{http://dx.doi.org/10.1103/PhysRevD.89.103012}{{\em Phys. Rev.}
  {\bfseries D89} no.~10, (2014) 103012},
\href{http://arxiv.org/abs/1402.5156}{{\ttfamily arXiv:1402.5156 [gr-qc]}}.

\bibitem{Ashton:2018jfp}
G.~Ashton {\em et~al.}, ``{BILBY: A user-friendly Bayesian inference library
  for gravitational-wave astronomy},''
  \href{http://dx.doi.org/10.3847/1538-4365/ab06fc}{{\em Astrophys. J. Suppl.}
  {\bfseries 241} no.~2, (2019) 27},
  \href{http://arxiv.org/abs/1811.02042}{{\ttfamily arXiv:1811.02042
  [astro-ph.IM]}}.

\bibitem{Romero-Shaw:2020owr}
I.~M. Romero-Shaw {\em et~al.}, ``{Bayesian inference for compact binary
  coalescences with bilby: validation and application to the first
  LIGO\textendash{}Virgo gravitational-wave transient catalogue},''
  \href{http://dx.doi.org/10.1093/mnras/staa2850}{{\em Mon. Not. Roy. Astron.
  Soc.} {\bfseries 499} no.~3, (2020) 3295--3319},
  \href{http://arxiv.org/abs/2006.00714}{{\ttfamily arXiv:2006.00714
  [astro-ph.IM]}}.

\bibitem{Dietrich:2017aum}
T.~Dietrich, S.~Bernuzzi, and W.~Tichy, ``{Closed-form tidal approximants for
  binary neutron star gravitational waveforms constructed from high-resolution
  numerical relativity simulations},''
  \href{http://dx.doi.org/10.1103/PhysRevD.96.121501}{{\em Phys. Rev. D}
  {\bfseries 96} no.~12, (2017) 121501},
  \href{http://arxiv.org/abs/1706.02969}{{\ttfamily arXiv:1706.02969 [gr-qc]}}.

\bibitem{Dietrich:2018uni}
T.~Dietrich {\em et~al.}, ``{Matter imprints in waveform models for neutron
  star binaries: Tidal and self-spin effects},''
  \href{http://dx.doi.org/10.1103/PhysRevD.99.024029}{{\em Phys. Rev. D}
  {\bfseries 99} no.~2, (2019) 024029},
  \href{http://arxiv.org/abs/1804.02235}{{\ttfamily arXiv:1804.02235 [gr-qc]}}.

\bibitem{Castro:2022mpw}
G.~Castro, L.~Gualtieri, A.~Maselli, and P.~Pani, ``{Impact and detectability
  of spin-tidal couplings in neutron star inspirals},''
  \href{http://arxiv.org/abs/2204.12510}{{\ttfamily arXiv:2204.12510 [gr-qc]}}.

\bibitem{Essick:2017wyl}
R.~Essick, S.~Vitale, and M.~Evans, ``{Frequency-dependent responses in third
  generation gravitational-wave detectors},''
  \href{http://dx.doi.org/10.1103/PhysRevD.96.084004}{{\em Phys. Rev. D}
  {\bfseries 96} no.~8, (2017) 084004},
  \href{http://arxiv.org/abs/1708.06843}{{\ttfamily arXiv:1708.06843 [gr-qc]}}.

\bibitem{LIGOScientific:2016wof}
{\bfseries LIGO Scientific} Collaboration, B.~P. Abbott {\em et~al.},
  ``{Exploring the Sensitivity of Next Generation Gravitational Wave
  Detectors},'' \href{http://dx.doi.org/10.1088/1361-6382/aa51f4}{{\em Class.
  Quant. Grav.} {\bfseries 34} no.~4, (2017) 044001},
  \href{http://arxiv.org/abs/1607.08697}{{\ttfamily arXiv:1607.08697
  [astro-ph.IM]}}.

\bibitem{Chatziioannou:2021tdi}
K.~Chatziioannou, ``{Uncertainty limits on neutron star radius measurements
  with gravitational waves},''
  \href{http://dx.doi.org/10.1103/PhysRevD.105.084021}{{\em Phys. Rev. D}
  {\bfseries 105} no.~8, (2022) 084021},
  \href{http://arxiv.org/abs/2108.12368}{{\ttfamily arXiv:2108.12368 [gr-qc]}}.

\bibitem{Gamba:2020wgg}
R.~Gamba, M.~Breschi, S.~Bernuzzi, M.~Agathos, and A.~Nagar, ``{Waveform
  systematics in the gravitational-wave inference of tidal parameters and
  equation of state from binary neutron star signals},''
  \href{http://dx.doi.org/10.1103/PhysRevD.103.124015}{{\em Phys. Rev. D}
  {\bfseries 103} no.~12, (2021) 124015},
  \href{http://arxiv.org/abs/2009.08467}{{\ttfamily arXiv:2009.08467 [gr-qc]}}.

\bibitem{Narikawa:2019xng}
T.~Narikawa, N.~Uchikata, K.~Kawaguchi, K.~Kiuchi, K.~Kyutoku, M.~Shibata, and
  H.~Tagoshi, ``{Reanalysis of the binary neutron star mergers GW170817 and
  GW190425 using numerical-relativity calibrated waveform models},''
  \href{http://dx.doi.org/10.1103/PhysRevResearch.2.043039}{{\em Phys. Rev.
  Res.} {\bfseries 2} no.~4, (2020) 043039},
  \href{http://arxiv.org/abs/1910.08971}{{\ttfamily arXiv:1910.08971 [gr-qc]}}.

\bibitem{Volkel:2022utc}
S.~H. V\"olkel and C.~J. Kr\"uger, ``{Constraining the Nuclear Equation of
  State from Rotating Neutron Stars},''
  \href{http://arxiv.org/abs/2203.05555}{{\ttfamily arXiv:2203.05555 [gr-qc]}}.

\bibitem{Tonetto:2021ovc}
L.~Tonetto, A.~Sabatucci, and O.~Benhar, ``{Impact of three-nucleon forces on
  gravitational wave emission from neutron stars},''
  \href{http://dx.doi.org/10.1103/PhysRevD.104.083034}{{\em Phys. Rev. D}
  {\bfseries 104} no.~8, (2021) 083034},
  \href{http://arxiv.org/abs/2106.16131}{{\ttfamily arXiv:2106.16131
  [nucl-th]}}.

\bibitem{Wijngaarden:2022sah}
M.~Wijngaarden, K.~Chatziioannou, A.~Bauswein, J.~A. Clark, and N.~J. Cornish,
  ``{Probing neutron stars with the full premerger and postmerger gravitational
  wave signal from binary coalescences},''
  \href{http://dx.doi.org/10.1103/PhysRevD.105.104019}{{\em Phys. Rev. D}
  {\bfseries 105} no.~10, (2022) 104019},
  \href{http://arxiv.org/abs/2202.09382}{{\ttfamily arXiv:2202.09382 [gr-qc]}}.

\bibitem{Breschi:2021tbm}
M.~Breschi, A.~Perego, S.~Bernuzzi, W.~Del~Pozzo, V.~Nedora, D.~Radice, and
  D.~Vescovi, ``{AT2017gfo: Bayesian inference and model selection of
  multicomponent kilonovae and constraints on the neutron star equation of
  state},'' \href{http://dx.doi.org/10.1093/mnras/stab1287}{{\em Mon. Not. Roy.
  Astron. Soc.} {\bfseries 505} no.~2, (2021) 1661--1677},
  \href{http://arxiv.org/abs/2101.01201}{{\ttfamily arXiv:2101.01201
  [astro-ph.HE]}}.

\end{thebibliography}\endgroup

\end{document}